\RequirePackage{fix-cm}

\documentclass[article]{aa}
\usepackage[utf8]{inputenc}
\usepackage{gensymb}
\usepackage[english]{babel}
\usepackage{appendix}
\usepackage{graphicx, subcaption}
\usepackage{hyperref}
\usepackage{makecell}

\hypersetup{
    colorlinks,
    linkcolor={red!50!black},
    citecolor={blue!50!black},
    urlcolor={blue!80!black}
}

\def\vecsign{\mathchar"017E}
\def\dvecsign{\smash{\stackon[-2.33pt]{\vecsign}{\rotatebox{180}{$\vecsign$}}}}
\def\dvec#1{\def\useanchorwidth{T}\stackon[-4.2pt]{#1}{\,\dvecsign}}
\usepackage{stackengine}
\stackMath

\usepackage{colortbl} %Allows the coloring of table cells
\usepackage{multirow}
\usepackage{xcolor}
\usepackage{soul}

\newcommand*{\genbf}[1]{\ifmmode\mathbf{#1}\else\textbf{#1}\fi}

\title{The influence of flux rope heating models on solar prominence formation}
\author{N. Brughmans\inst{\ref{KUL}} \and J. M. Jenkins\inst{\ref{KUL}} \and R. Keppens\inst{\ref{KUL}}}
\institute{Centre for mathematical Plasma-Astrophysics, Celestijnenlaan 200B, 3001 Leuven, KU Leuven, Belgium \\\email{nicolas.brughmans@kuleuven.be} \label{KUL}}
\date{Received 2022 / Accepted 2022}

\abstract
{Prominences are cool, dense clouds suspended within the solar corona. Their in situ formation through the levitation-condensation mechanism is a textbook example of the thermal instability, where a slight energy imbalance leads to a runaway process resulting in condensed filamentary structures embedded within the concave-up portions of a flux rope. The detailed interplay between local radiative losses and the global heating of the solar corona is investigated here for prominence-forming flux rope structures.}
{We begin by exploring the influence of two classes of commonly adopted heating models on the formation behaviour of solar prominences. These models consider either an exponential variation dependent on height alone, or local density and magnetic field conditions. We highlight and address some of the limitations inherent to these early approximations by proposing a new, dynamic 2D flux rope heating model that qualitatively accounts for the 3D topology of the twisted flux rope field.}
{We performed 2.5D grid-adaptive numerical simulations of prominence formation via the levitation-condensation mechanism. A linear force-free arcade is subjected to shearing and converging motions, leading to the formation of a flux rope containing material that may succumb to thermal instability. The eventual formation and subsequent evolution of prominence condensations was then quantified as a function of the specific background heating prescription adopted. For the simulations that consider the topology of the flux rope, reduced heating was considered within a dynamically evolving ellipse that traces the flux rope cross-section. This ellipse is centred on the flux rope axis and tracked during runtime using an approach based on the instantaneous magnetic field curvature.}
{We find that the nature of the heating model is clearly imprinted on the evolution and morphology of any resulting prominences: one large, low-altitude condensation is obtained for the heating model based on local parameters, while the exponential model leads to the additional formation of smaller blobs throughout the flux rope which then relocate as they tend towards achieving hydrostatic equilibrium. Finally, a study of the condensation process in phase space reveals a non-isobaric evolution with an eventual recovery of uniform pressure balance along flux surfaces.} {}

\begin{document}

\maketitle

% Introduction section

\section{Introduction} \label{sec:intro}
%\textit{Research question: what is influence of different heating models on prominence formation?}
%\textit{Why is problem important: arbitrary choice of heating model, towards increased realism}
%\textit{What lacks in current understanding: there is no definitive heating model yet, how can we increase realism and how do simulations change?}

% \subsection{Prominence formation}

Prominences are cool, dense plasma structures in the solar corona, suspended by the magnetic field. They are usually found above polarity inversion lines (PILs) -- features found in magnetogram observations where the normal component of the photospheric magnetic field changes sign \citep[][]{Mackay:2010,Vial2015,Gibson:2018}. In recent years, many high resolution simulations have been performed, both in 2D and 3D, that have investigated prominence formation through various mechanisms. These mechanisms usually require the magnetic field topology local to the condensations to be concave upwards with respect to gravity; levitation against gravity is facilitated by means of the magnetic Lorentz force (pressure and tension), as first shown by the self-similar model of \cite{kippenhahn1957theorie}. Generally speaking, two main formation processes are discerned: (i) evaporation-condensation, where localised footpoint heating of a magnetic arcade produces upflows which collect in any pre-existing dipped fields \citep{Xia2012,Xia2014,Xia2016a} -- a theory shared with the phenomenon of coronal rain \citep{Li2021}; and (ii) levitation-condensation \citep{KY2015,KY2018,jack2020} or its variant reconnection-condensation \citep{KY2017}, where a flux rope is dynamically formed through shearing and converging motions at lower altitudes -- a sheared magnetic arcade, whose footpoints are subject to those motions, eventually changes internal connectivity via magnetic reconnection and builds a twisted, helical field \citep{ballegooijen_martens}. In doing so, the reconnection may redistribute or scoop up material from the chromosphere or lower corona, which remains suspended by the flux rope (levitation). The condensation aspect of all three formation mechanisms relies on a runaway cooling effect triggered by perturbations to the energy loss rate, in other words density and temperature variations, known as the thermal instability \citep{parker, field}.

The thermal instability is of paramount importance in many fields of astrophysics. The process leads to the formation of filamentary structures, providing an excellent alternative to the gravitational instability in the interstellar medium \citep{interstellar_filament} or galaxy clusters \citep{galaxy_cluster_filament}. Numerical implementations commonly assume a purely optically thin ambient medium, such as is the case for the solar corona, where energy transported as radiation is able to `free-stream' outwards. Since perturbations to density and temperature can lead to local increases in these radiative losses, variations feed back into the energy balance so as to yet further increase the density and decrease the temperature. The criteria that govern the (in)stability of this delicate equilibrium were first formulated by \cite{parker} and fully established through the seminal work of \cite{field}. The latter author obtained an analytic dispersion relation in an idealised setting that governs the linear behaviour of the mode. Recently, \cite{ClaesN2019Tsom} and \cite{ClaesN2020TiFa} revisited these results and numerically studied a local coronal volume pervaded by interacting slow magneto-acoustic waves in 2D and 3D settings, respectively. The authors showed that the thermal mode eventually takes over, with flows along the magnetic field that drive material towards a thermally unstable location, leading to the formation of a filamentary structure initially oriented almost perpendicular to the magnetic field. Subsequent small differences in ram pressure from both sides force the redistribution of these structures along the magnetic field; in 3D, an intriguing misalignment was found between the condensations and the magnetic field. How different cooling curves impact the above process was investigated by \cite{joris2021}, who introduced a bootstrap measure such that extremely thin filaments could be handled at unprecedented numerical resolutions. Further evidence for the importance of thermal instability within the solar atmosphere was recently provided by \cite{Claes2021}, who examined the full magnetohydrodynamic (MHD) spectrum of a realistic stratified and magnetised atmosphere. The study highlighted that much of the solar chromosphere, transition region and corona can be liable to unstable or overstable magneto-thermal modes.

% Under the right conditions, a decrease in temperature will lead to even larger energy losses, driving the instability. Since the radiative losses depend on $\rho^2$ (see below), an increasingly dense condensation will continue to grow even if the radiative losses start to decrease for lower temperatures. The condensation will keep cooling down and become heavier, until a minimum temperature is reached, at which point much of the condensed plasma is neutral due to recombination and the assumption of an optically thin medium breaks down.

% Prominence oscillations? Veronika, link to Xia2014,2016 \citep{Jercic2022}. Could also refer to the work of the canary islands group, perhaps Valeriia.
% Post-flare loops? Wenzhi's paper(s?) on flares, since these incorporate the same kind of setup but with a focus on a smaller part of the domain, i.e. the reconnection point below the flux rope and the flare loop that it creates: post-flare coronal rain \citep{Ruan2021}

The recent study of \citet{jack2020} considered the formation and evolution of 2.5D flux rope\,--\,prominence systems via the thermal instability as a component of the common levitation-condensation mechanism \citep{KY2015}. Their solar coronal model adopted an exponential background heating profile meant to balance the local radiative and conductive losses of the system governed mechanically by hydrostatic equilibrium \citep{Fang2013}. Once the flux rope\,--\,prominence system formed, the simulation yielded high-resolution demonstrations of the dynamic formation of prominence condensations via the thermal instability. The authors clarified how the fully non-linear evolution can be related to sequential linear MHD mode stability considerations, with the thermal mode driving the condensation process. In their 2.5D simulations, the field-aligned thermal conduction term within their energy equation in essence thermally isolated the flux rope plasma from the background corona, but the time-independent, exponential background heating component still supplied a small but non-negligible contribution to this plasma at all times. This directly influences the thermal balance and in turn artificially lengthens the formation timescales of thermal instability \citep{joris2021}. This aspect is of critical importance when considering the dynamic formation, evolution, and decay processes in prominences, for instance the timescales of prominence mass recycling, as recently highlighted by the pioneering work of \cite{KY2018}. 
It should be stated that thermal instability in the context of solar prominence formation also depends on the presence of the chromosphere and transition region \citep[][]{Klimchuk:2019b}. Through thermal conduction, much of the energy of the solar corona is transported downwards, which can provide a more efficient mechanism for energy loss than through radiative losses alone. Moreover, the inclusion of the extra mass reservoir opens the aforementioned evaporation-condensation evolutionary pathways to modifying the eventual properties of the prominence material.

% \subsection{Coronal heating}

The thermal instability relies on the intricate balance between energy loss, gain, and transport mechanisms. Within the solar corona this balance is generally regulated by ambient heating, downwards thermal conduction and optically thin radiative cooling, with additional considerations for  compression and Joule heating, for instance. However, as the cause of the $1$~MK solar corona remains a long-standing open problem, often ad hoc background heating models are artificially imposed. Two classes of artificial heating models are generally discerned \citep{Mandrini}. The first class describes the heating rate as a function of height in the corona based on observations or theoretical models, typically formulated assuming an exponential decrease with height as in \cite{jack2020} (see, e.g. \cite{exponential_heating}, for the observational basis, but also \cite{Fang2013}, \cite{Xia2016a}, \cite{Zhao2017} and \cite{Fan2017} for similar numerical implementations). The second class considers physically or observationally derived scaling laws inside coronal loops \citep[e.g.][]{dahlburg2018}, but the models can be applied more generally to the solar atmosphere. These scaling laws provide a parameterised heating rate based on local plasma conditions,
\begin{equation} \label{eq:heating_general}
	\mathcal{H} \sim B^\alpha \rho^\beta L^\gamma,
\end{equation}
where $B$ is the magnetic field strength, $\rho$ the density, and $L$ the length of the field line through the local volume. This heating model is characterised by the combination of powers $(\alpha,\beta,\gamma)$. A comprehensive summary of several known possible mechanisms behind these scaling laws is given in Table~5 of \cite{Mandrini}, ranging from reconnection (DC models) to wave dissipation (AC models), with dependencies that sometimes include additional parameters beyond the three given above. 

Each of these two classes (exponential or parameterised through Eq.~(\ref{eq:heating_general})) of heating prescriptions has been successfully employed in simulations of both coronal loops \citep{Mok2008, Mok2016} and prominences \citep{KY2015, KY2017, KY2018, Xia2012, Xia2014}. Incorporating field line length dependence is computationally demanding, as it requires tracing field lines for every cell in the simulation domain during runtime. This is non-trivial to realise in a grid-adaptive, parallelised framework like MPI-AMRVAC \citep{AMRVAC}, although recent work addressing solar flares with electron beam ingredients along evolving reconnecting field lines used this in \cite{Ruan2021}. \cite{Mok2016} provide an alternative in the context of coronal loop simulations through approximating the field line length $L$ by the radius of curvature of the magnetic field $R$, which is the inverse of the magnetic curvature $\kappa$. This first order approximation is valid when the loops are almost circular. In the context of prominences, however, this approach proves unable to capture the complex twisted field that comprises a flux rope. 

In this work, we implement a wide array of heating prescriptions and compare their influence on prominence formation through the 2.5D levitation-condensation models of \cite{jack2020}. For Eq.~\ref{eq:heating_general}, a comparison between two such heating models, $(\alpha,\beta)=(2,0)$ and $(\alpha,\beta)=(0,1)$, with $\gamma=0$, was previously presented by \cite{KY2015} as they introduced the levitation-condensation model. They found that for those heating models that depend on magnetic field strength, condensations appear only when the flux rope is formed through anti-shearing motions, where the arcade footpoint motions are directed so as to decrease the magnetic shear of the initial condition. We extend these results here by looking at the morphology and physical properties of the formed prominences, doing so for each of the aforementioned heating prescriptions and senses of shear, in addition to a new `reduced heating' model that furthermore approximates the influence of the $L$ parameter for flux ropes. Moreover, a phase space distribution of the condensation process reveals a non-isobaric evolution of thermal instability, where constant pressure is eventually reached along the flux surfaces that contain the prominence.

In Section~\ref{sec:methods}, we detail the simulation setup used throughout this work, together with the heating models that we considered and our method for flux rope tracking. In Section~\ref{sec:results}, we list the results from our simulations, which we further discuss in Section~\ref{sec:discussion}.

% Methods section

\section{Numerical setup and equations} \label{sec:methods}
%\textit{See thesis and add the necessary stuff (e.g., on curvature)}
%\textit{Equations: MHD with resistivity, radiative cooling and heating}
%\textit{Numerical: description of the simulation domain, boundaries, numerical methods to treat physics. Refer to JK2020 when nothing changed.}
%\textit{Setup: atmosphere in thermal and equilibrium, driving motions (refer to JK2020)}

% This might belong more in the Simulation setup part.
As already indicated, solar prominences are phenomena found embedded within the solar corona. Following the original levitation-condensation setup of \cite{KY2015}, we restrict our region of interest to the coronal volume and neglect the chromosphere below. We consider a 2D simulation domain in the $x-y$-plane (horizontal-vertical), with vector components in the invariant $z$-direction (horizontal) to complete the 2.5D description. To carry out the simulation, we make use of the fully open-source, adaptive-grid, parallelised MPI-AMRVAC toolkit\footnote{\hyperlink{http://amrvac.org}{http://amrvac.org}} \citep{AMRVAC}. We simulate a cross-section of $24\times 25$~Mm with $x \in [-12,12]$~Mm and $y \in [0,25]$~Mm, taking a base resolution of $96\times 96$ cells with three additional levels of Adaptive Mesh Refinement (AMR). This yields a maximum resolution of approximately $24$~Mm~$/~(96\times 2^3) = 31.25$~km, which corresponds to the lower-resolution simulations of \cite{jack2020}. The AMR criteria are based on sharp gradients in the local density and the magnetic field components following the Löhner prescription \citep{lohnerAMR}.

\subsection{Equations and physics} \label{sec:equations_and_physics}

The full set of MHD equations solved by MPI-AMRVAC, including all non-ideal or non-adiabatic effects and source terms, are equivalent to (the code actually solves the related system using conservative variables),
\begin{align} 
	 \frac{\partial\rho}{\partial t} + \nabla\cdot(\rho\mathbf{v}) &= 0, \label{eq:MHD_full_rho} \\[.7ex] 
	 \rho\frac{\partial\mathbf{v}}{\partial t} + \rho\mathbf{v}\cdot\nabla\mathbf{v} + \nabla p - \mathbf{j}\times\mathbf{B} - \rho\mathbf{g} &= 0, \label{eq:MHD_full_momentum} \\[.7ex]
	 \frac{\partial\mathbf{B}}{\partial t} - \nabla\times(\mathbf{v}\times\mathbf{B}) + \nabla\times\eta\mathbf{j} &= 0, \label{eq:MHD_full_magnetic}
\end{align}
\begin{align}
	 \begin{split} \rho\frac{\partial T}{\partial t} + \rho\mathbf{v}\cdot\nabla T \\
	 	+ (\gamma-1)\left[ p\nabla\cdot\mathbf{v} + \rho\mathcal{L} - \eta |\mathbf{j}|^2 - \nabla\cdot(\dvec \kappa\cdot\nabla T) \right] &= 0, \end{split} \label{eq:MHD_full_temperature}
\end{align} 
as given in \cite{goedbloed_keppens_poedts_2019}\footnote{Correcting on a typo in Eq.~3 of \citet{jack2020}.}. The thermodynamic variables are related through an equation of state, for which we use the ideal gas law,
\begin{equation} \label{eq:ideal_gas_law}
	p\mu = \mathcal{R} \rho T.
\end{equation}
Here, $\mu$ is the mean molecular mass in proton masses $m_p$ and $\mathcal{R} \approx 3.81$ J K$^{-1}$ is the ideal gas constant. We assume a monatomic gas so that the ratio of specific heats $\gamma = 5/3$. Governing the magnetic field are Gauss' law for magnetism and Ampère's law,
\begin{equation} \label{eq:maxwell}
	 \nabla\cdot \mathbf{B} = 0, \quad \nabla\times \mathbf{B} = \mu_0 \mathbf{j},
\end{equation}
where 
%$\mu_0 = 1.26\times 10^{-6}$ N A$^{-2}$
$\mu_0 = 4\pi$ is the permeability of vacuum. The (dependent) variables appearing in this closed system of equations, are the density $\rho$, velocity $\mathbf{v}$, gas pressure $p$, current density $\mathbf{j}$, magnetic field $\mathbf{B}$ and temperature $T$.

The temperature equation \eqref{eq:MHD_full_temperature} contains energy sinks and sources within the square brackets. Firstly, the heat-loss function $\rho\mathcal{L} = n_\textrm{H} n_\textrm{He} \Lambda(T) - \mathcal{H}$ contains the net volumetric energy loss from optically thin radiative cooling  $n_\textrm{H} n_\textrm{He} \Lambda(T)$ and heating $\mathcal{H}$, with units ergs~cm$^{-3}$~s$^{-1}$ \citep{field}. We note that the cooling rate depends on the hydrogen and helium number densities $n_\textrm{H,He}$ and temperature $T$ through a cooling curve $\Lambda$, constructed using tabulated energy losses. The volumetric heating rate $\mathcal{H}$ -- although currently unknown -- must necessarily depend on a number of quantities, such as density $\rho$ or magnetic field strength $B$, and we consider a few possibilities throughout this work \citep[][]{Mandrini}. Secondly, Ohmic heating $\sim \eta |\mathbf{j}|^2$ is included in line with the consideration of magnetic resistivity. The coefficient $\eta$ has an artificially high value of $2.33\times10^8$~m$^2$~s$^{-1}$($= 0.002$ in code units), ensuring that we dominate any possible numerical diffusion, and allowing the reconnection necessary during the early formation phase of the flux rope. Thirdly, anisotropic thermal conduction $\nabla\cdot(\dvec \kappa\cdot\nabla T)$ is included, where for the parallel component we take the Spitzer conductivity $\kappa_\parallel~=~8\times~10^{-7}~T^{5/2}$~ergs~cm$^{-1}$~s$^{-1}$~K$^{-1}$ \citep{spitzer2006}. Due to being eight orders of magnitude smaller than the parallel component under coronal conditions, perpendicular thermal conduction is neglected throughout this study.

The plasma is assumed to be completely ionised \citep[an assumption that may be invalid inside cool and dense prominences][]{hillier2019ion,braileanu2021one,braileanu2021two}, and is partly composed of helium ions with assumed abundance $n_\textrm{He}  = 0.1 n_\textrm{H}$. The mean molecular mass is hence,
\begin{equation} \label{eq:mean_molecular_mass}
	\mu = \frac{\rho / m_p}{n} = \frac{n_\textrm{H}+4n_\textrm{He}}{ 2n_\textrm{H}+3n_\textrm{He}} \approx 0.6.
\end{equation}

For consistency with earlier work \citep{jack2020}, we adopt a non-constant gravitational acceleration,
\begin{equation} \label{eq:gy}
    \mathbf{g}(y) = - g_\textrm{cor} \frac{R_\sun^2}{(R_\sun+y)^2} \mathbf{\hat{y}},
\end{equation}
with $y$ the height above the solar surface and $g_\textrm{cor} = 274$~m~s$^{-2}$ the gravitational acceleration at the solar surface, located at radius $R_\sun = 695\,700$~km. In our simulation domain, the gravitational acceleration falls from $g_\textrm{cor}$ at the bottom to $0.93 g_\textrm{cor}$ at the top ($y=25$ Mm).

The MHD equations \eqref{eq:MHD_full_rho}$-$\eqref{eq:MHD_full_magnetic} are solved using a threestep, third-order Runge-Kutta method with the HLL flux scheme \citep{HLL}. To limit spurious oscillations, we use the third-order asymmetric \texttt{CADA3} slope limiter proposed by \cite{CADA3}. As a divergence fix for the magnetic field, we use the \texttt{linde} method to perform parabolic cleaning \citep{keppens2003} and reduce monopole errors through the bottom boundary driving where a second order central differences approximation for $\nabla\cdot\mathbf{B} = 0$ is implemented. MPI-AMRVAC allows for a split treatment of the magnetic field, fixing a steady background field $\mathbf{B}_0$ while performing calculations for the perturbed field $\mathbf{B}_1$ as described by \cite{tanaka1994}, which is also possible for force-free fields \citep{AMRVAC1}. Magnetic field splitting ensures accuracy and efficiency by better capturing behaviour at low plasma-beta. We also solve an auxiliary energy equation for the internal energy density so as to handle typical coronal low-beta regions and replace faulty pressure values when necessary. For the radiative cooling physics, we use the \texttt{Colgan\_DM} cooling curve by \cite{ColganJ2008RLoS}, modified for lower temperatures by \cite{DM}, which is suitable for this study \citep{joris2021}. The cooling curve is interpolated at $12\,000$ points and calculated at runtime using the exact integration method \citep{exact}. A minimal allowed temperature of $10^3$~K is enforced throughout the simulations, well below the typical minimum temperatures found in our simulations ($\sim 7000$~K). MPI-AMRVAC uses normalisation of the plasma variables, which is fixed by the following choice: number density $\Tilde{n}_{\textrm{H}} = 10^8$~cm$^{-3}$, length $\Tilde{L} = 10^8$~cm and temperature $\Tilde{T} = 10^6$~K.

\subsection{Simulation setup} \label{sec:simulation_setup}
%\textit{Hydrostatic equilibrium, initial LFF magnetic field, heating models in thermal equilibrium}

As initial conditions, we set up a stratified solar coronal atmosphere in hydrostatic and thermal near-equilibrium. Assuming an isothermal corona at a temperature of $T_0 = 1$~MK and assuming a plane-parallel atmospheric stratification, the momentum equation \eqref{eq:MHD_full_momentum} reduces to a separable ODE if the magnetic field is force-free. % as will be specified below.
%The equation reduces to
%\begin{equation}\label{eq:hydrostatic_pressure_diff_eq}
%	\frac{dp}{dy} =-\frac{\mu}{\mathcal{R}} \frac{p(y)}{T_0} g(y),
%\end{equation}
%where $\mu, \mathcal{R}$ are the constants from the ideal gas law \eqref{eq:ideal_gas_law}. 
The resulting equation is solved numerically with the trapezoid rule in our initial conditions, which is numerically convenient and equivalent to setting the analytic hydrostatic solution. In hydrostatic equilibrium, it follows from the ideal gas law \eqref{eq:ideal_gas_law} that pressure and density feature the same exponential variation with height,
\begin{equation}\label{eq:hydrostatic_profiles}
% 	p_0(y) &= p_\sun \exp\left( -\frac{y}{H(y)} \right), \quad \rho_0(y) = \rho_\sun \exp\left( -\frac{y}{H(y)} \right),\\
	p_0(y) = p_\textrm{cor} \exp\left( -\frac{y}{H(y)} \right), \quad \rho_0(y) = \rho_\textrm{cor} \exp\left( -\frac{y}{H(y)} \right),
\end{equation}
where,
\begin{equation} \label{eq:scale_height}
	H(y) = H_0 \frac{R_\sun + y}{R_\sun}, \quad \text{with } H_0 = \frac{\mathcal{R} T_0}{\mu g_\textrm{cor}},
\end{equation}
is a local scale height based on the hydrostatic pressure scale height $H_0 \approx 50$ Mm and $p_\textrm{cor} = 0.434$ dyn cm$^{-2}$ and $\rho_\textrm{cor} = 3.2\times 10^{-15}$ g cm$^{-3}$ are the pressure and density values at the base of the corona, respectively.

%At the top of the domain, $H(y) = 1.04 H_0$ and pressure and density are $2\%$ larger than if the gravitational acceleration $g_0$ were constant over the domain. We note the difference with \cite{KY2018}, where $H_0 \approx 30$ Mm since the authors assume $\mu = 1$. Indeed, we have $30 / 0.6 = 50$ Mm since $\mu = 0.6$.

The magnetic configuration within the simulation starts with the linear force-free sheared magnetic arcade $\mathbf{B}_0$ given by \cite{KY2015,KY2017,KY2018} and \cite{jack2020}. Then the magnetic field strength varies as,
% Left out the arcade prescription.
% \begin{align} 
%     B_{0x} &= -\left(\frac{2L_a}{\pi a}\right) B_a \cos\left(\frac{\pi x}{2L_a}\right) \exp\left(\frac{-y}{H(y)}\right), \label{eq:Bx_new} \\
%     B_{0y} &= B_a \sin\left(\frac{\pi x}{2L_a}\right) \exp\left(\frac{-y}{H(y)}\right), \label{eq:By_new} \\
%     B_{0z} &= \sqrt{1-\left(\frac{2L_a}{\pi a}\right)^2} B_a \cos\left(\frac{\pi x}{2L_a}\right) \exp\left(\frac{-y}{H(y)}\right) \label{eq:Bz_new}, \\
% \end{align}
\begin{equation} \label{eq:magnetic_field_strength_new}
		B_0(y) = B_a \exp\left(\frac{-y}{H_0}\right).
\end{equation}
Here, $B_a = 10$ G is the field strength at the bottom of the domain.
% and $a = H_0$ is the magnetic scale height, chosen equal to the pressure scale height for convenience later on (Sect.~\ref{sec:heating_exponential_mixed}). 
% The parameter $L_a = 24$ Mm controls the width of 
The magnetic arcade has exactly one half period in the simulation domain with the PIL at $x=0$ Mm, setting an initial shear angle of $\approx$~9$\degree$ with the out-of-plane normal. Our choice of $B_a$ corresponds to the strongest magnetic field considered by \cite{jack2020}. The field strength, gas pressure and density share approximately the same height profile such that magnetic pressure $\left( \sim B^2 \right)$ decreases twice as fast. The initial sheared arcade prescription is taken to be the steady background field $\mathbf{B}_0$ used in the aforementioned magnetic field splitting. An image of this initial setup can be found in \cite[Fig. 1]{jack2020}.

% With these choices for $\rho_0$ and $\mathbf{B}_0$, the atmosphere is initially in mechanical equilibrium. Thermal equilibrium can be obtained if the local radiative losses are initially exactly balanced out by the background heating. As we prescribe and vary this background heating, sometimes not in perfect balance with the local losses, and as we have (numerical and explicit) resistivity, where any developing temperature gradient leads to thermal conduction, we first perform initial relaxations of a static atmosphere to end up with near-steady conditions and flows well below $10$~km~s$^{-1}$, the typical velocity of background flows observed in the solar atmosphere  \citep[e.g.][]{hassler1999}. After approximately $1.5$~h, the atmosphere settles in a near-equilibrium state independent of the heating model used so long as our $t=0$~s conditions had thermal losses and sinks of similar magnitude initially. Such relaxation of $1.5$~h is very long compared to the relevant dynamical time scales $\tau_A \approx 54$~s, $\tau_S \approx 266$~s determined by the Alfvén speed and the sound speed, respectively. Hence, for the actual simulations, the initial state is relaxed for a shorter period of approximately $1300$~s ($\approx 5 \times \tau_S$) before we drive the evolution within the bottom boundary, consistent with \cite{jack2020}. 

% Could stick this behind equations and physics?
\subsection{Boundary conditions} \label{sec:boundary_conditions}

We take the same boundary conditions described in \cite{jack2020}, where there is no outflow out of the domain and the magnetic field is vertical (in line with the periodicity) at the left and right edges. At the lower and upper boundary, the magnetic field is extrapolated.
To form the flux rope, we impose the shearing and converging motions at the bottom boundary as originally described in \cite{KY2015}. 
Following these authors, there is a distinction to be made between shearing and anti-shearing motions, which increase and decrease the magnitude of the magnetic field component in the invariant direction, respectively. The motion is prescribed as,
\begin{equation} \label{eq:driven_boundary_space_z}
    V_z(x,t) = \begin{cases}
         -V_x(x,t), & \text{shearing,} \\
         V_x(x,t), & \text{anti-shearing},
    \end{cases}
\end{equation}
where $V_x(x,t)$ is prescribed following Equations $13\,-\,17$ of \citet{jack2020}; the associated 12~km~s$^{-1}$ magnitude for driving and shearing motions is then far below the characteristic coronal Alfv\'en speed of $\approx$~442~km~s$^{-1}$. We adopt the driving prescription in time and space from this work but do not redefine the moment they are initiated to $t=0$. The driving hence begins at $t=1300$~s and lasts until $t=2800$~s.

\subsection{Heating models} \label{sec:heating_models}

We now include an overview of the heating models adopted in this work, and describe how detailed balance can be achieved against the initial radiative losses. Since these initial losses scale as $\rho_0(y)^2 \Lambda(T_0)$, they are explicitly known at $t=0$~s using Eq.~\eqref{eq:hydrostatic_profiles} and $T_0 = 1$ MK. For the required initial thermal equilibrium, any heating model has to fulfill the condition that there are no initial net energy losses or gains:
\begin{equation} \label{eq:thermal_equilibrium}
	0 =  \rho_0 \mathcal{L}_0 = \rho_0^2 \Lambda(T_0) - \mathcal{H}_0.
\end{equation}

%\subsubsection{Exponential heating model} \label{sec:heating_exponential}
\subsubsection{Exponential and mixed heating models} \label{sec:heating_exponential_mixed}

The first heating model is time independent and is hence completely determined by condition \eqref{eq:thermal_equilibrium}, as in for instance \cite{ClaesN2019Tsom} and \cite{joris2021}. The heating rate then falls exponentially with height, since it is the absolute value of the initial cooling rate. We henceforth refer to this model as the exponential heating model, or model \texttt{E}. The heating prescription, which has been successfully implemented by many previous authors \citep{Fang2013,Xia2016a,Zhao2017,jack2020} then becomes:
\begin{equation} \label{eq:heating_exponential}
	\mathcal{H} = \rho_{cor}^2 \Lambda(T_0)  \exp\left( -\frac{2y}{H(y)} \right).
\end{equation}
%We note that we have defined a constant heating rate per unit volume $\mathcal{H}$. Choosing a constant heating rate per unit mass $h$ such that $\rho\mathcal{L} = -\rho^2\Lambda(T) + \rho h$ gives exactly the same heating model at $t=0$, but as time progresses, the heating rate will really depend on the \textit{local} plasma density. An example of this heating prescription in literature is the second heating model used in \citep{KY2015} but it is part of a bigger class of heating models presented here.

%\subsubsection{Mixed heating models} \label{sec:heating_mixed}
A second class of heating models is based on those heating scaling laws deduced for coronal loops. The mixed heating model, or model \texttt{M}, is of the general form,
\begin{equation}\label{eq:heating_mixed_general}
	\mathcal{H} = c B^\alpha \rho^\beta,
\end{equation}
where $(\alpha, \beta) \in \mathbb{R}^2$ and $c(y)$ is a function determined by the equilibrium condition \eqref{eq:thermal_equilibrium}. Instead of using Eq.~\eqref{eq:magnetic_field_strength_new} for the initial magnetic field strength, we approximate it with the exponential profile with varying scale height for convenience in the expression below. This sets up an approximate initial equilibrium which relaxes during the first $1300$~s, with the initial background heating given by,
\begin{equation}
	\mathcal{H}_0 = c(y) B_a^\alpha \exp\left( -\frac{\alpha y}{H(y)} \right) \rho_{cor}^\beta \exp\left( -\frac{\beta y}{H(y)} \right).
\end{equation}
The function $c(y)$ can then be found as,
\begin{equation}\label{eq:heating_mixed_constant}
	c(y) = \frac{\rho_{cor}^{2-\beta}}{B_a^\alpha} \Lambda(T_0)  \exp\left( -\frac{(2-\alpha-\beta)y}{H(y)} \right),
\end{equation}
where the adopted approximate magnetic scale height allows the exponentials to be combined. We note that choosing $\alpha+\beta=2$ makes $c$ a constant. This is preferable both from a theoretical and practical point of view: if $\alpha+\beta\neq 2$ then there is still a steady, exponential component to the heating, losing the truly local nature of the heating model and maintaining the potentially unphysical residual heating to the flux rope. Hence, all the mixed heating models considered in this work are chosen to satisfy $\alpha+\beta=2$, as in the model $(\alpha,\beta) = (2.5,-0.5)$ used by \cite{Mok2008}.

Indeed, heating prescriptions depending on local parameters have been used before, most notably by \cite{Mok2008} and \cite{KY2017} in its two-parameter form above. Several other combinations of parameters $(\alpha,\beta)$ have been used throughout the literature, most often with $\beta = 0$ so as to obtain a heating rate that depends purely on the local magnetic energy density, hereafter model \texttt{M0} (magnetic heating) \citep{KY2015, KY2018, Xia2012, Xia2014}. We investigate the influence of models using a small, but increasingly, negative $\beta$ on the prominence formation, which we term models \texttt{M1} (quasi-magnetic heating, $\beta=-0.1$), \texttt{M2} ($\beta=-0.2)$, and \texttt{M3} \citep[][$\beta=-0.5$]{Mok2008} -- an overview of the models used in this work and their prescription is given in Table~\ref{tab:heating_comparison}. With the inverse density dependence, the heating prescription is meant to be magnetic energy-based, but with an additional `penalty' for regions of increased density. When performing prominence simulations, this is a desirable property since then the heating rate will decrease in any locations subject to a density enhancement. We note that the previous exponential heating model \texttt{E} is equivalent to the mixed heating model with choice $(\alpha,\beta) = (0,0)$.

% Table has been incorporated in Table 2.

%\begin{table}
%	\centering
%	\begin{tabular}{ccc}
%		\hline\hline
%			 									& $\alpha$ & $\beta$ \\ \hline
%			Exponential (\texttt{E}) & $0.0$ &  $0.0$ \\
%			Magnetic (\texttt{M0}) & $2.0$ & $0.0$  \\
%			Quasi-magnetic (\texttt{M1}) & $2.1$ &  $-0.1$ \\
%			\texttt{M2} &  $2.2$ & $-0.2$ \\
%			\cite{Mok2008} (\texttt{M3}) & $2.5$ & $-0.5$ \\ \hline
%	\end{tabular}
%	\caption{Combinations of the powers $\alpha$ and $\beta$ used in the prominence simulations with mixed heating ($\mathcal{H}\sim B^\alpha \rho^\beta$). Each heating model has its own abbreviation for further reference. COULD MERGE THIS WITH TABLE CONTAINING RESULTS Y/N.}
%	\label{tab:heating_mixed_combos}
%\end{table}

\subsubsection{Reduced heating model} \label{sec:heating_reduced}

Finally, we propose a new heating model pertaining to 2.5D prominence formation simulations that effectively approximates the true 3D nature of a flux rope. We want to account for the ignored variation in $z$, and note that a flux rope within the solar atmosphere is typically characterised by extended field lines of some hundred Mm. \cite{KY2017} demonstrated that inside a flux rope, longer field lines preferentially cool down and host condensations notwithstanding a heating prescription that did not explicitly trace field lines. By assuming the action of heating instead scales exponentially along field lines rather than simply with height \citep[many 1D hydro models of prominence formation exist which inherently support the relevance of the field line length to the condensation process; see, e.g.][] {karpen2008condensation,Klimchuk:2019a,Pelouze:2022}, we may argue that the flux rope should experience a reduced heating rate in its central portion compared to the footpoint regions. In the 2.5D MHD scenario studied here, we thereby want to incorporate how field lines in 3D have varying lengths throughout our flux rope cross-section.

To this aim, we apply a time-dependent heating reduction that follows the flux rope cross-sectional shape throughout the simulations. This requires a dynamic detection of the flux rope during runtime, for which we propose a method that relies on the 2.5D local magnetic field curvature (i.e. curvature defined using the 3D magnetic field vector but omitting gradients in the $z$-direction). Such a local prescription removes the need for tracing the global field associated with the flux rope. This resulting reduced heating model is applied to both the exponential heating model (now termed \texttt{RE}) and the mixed model (\texttt{RM}) so as to approximate the thus far ignored dependence of the general heating model \eqref{eq:heating_general} on field line length $L$ inside the flux rope. 

% Assuming that our 2D simulation domain encompasses a vertical cross-section through the (horizontal) middle of the flux rope, we apply a reduced heating rate inside the detected flux rope compared to the surrounding corona.

\begin{figure}
	\centering
	\resizebox{\hsize}{!}{\includegraphics[trim={0 10 0 0 },clip]{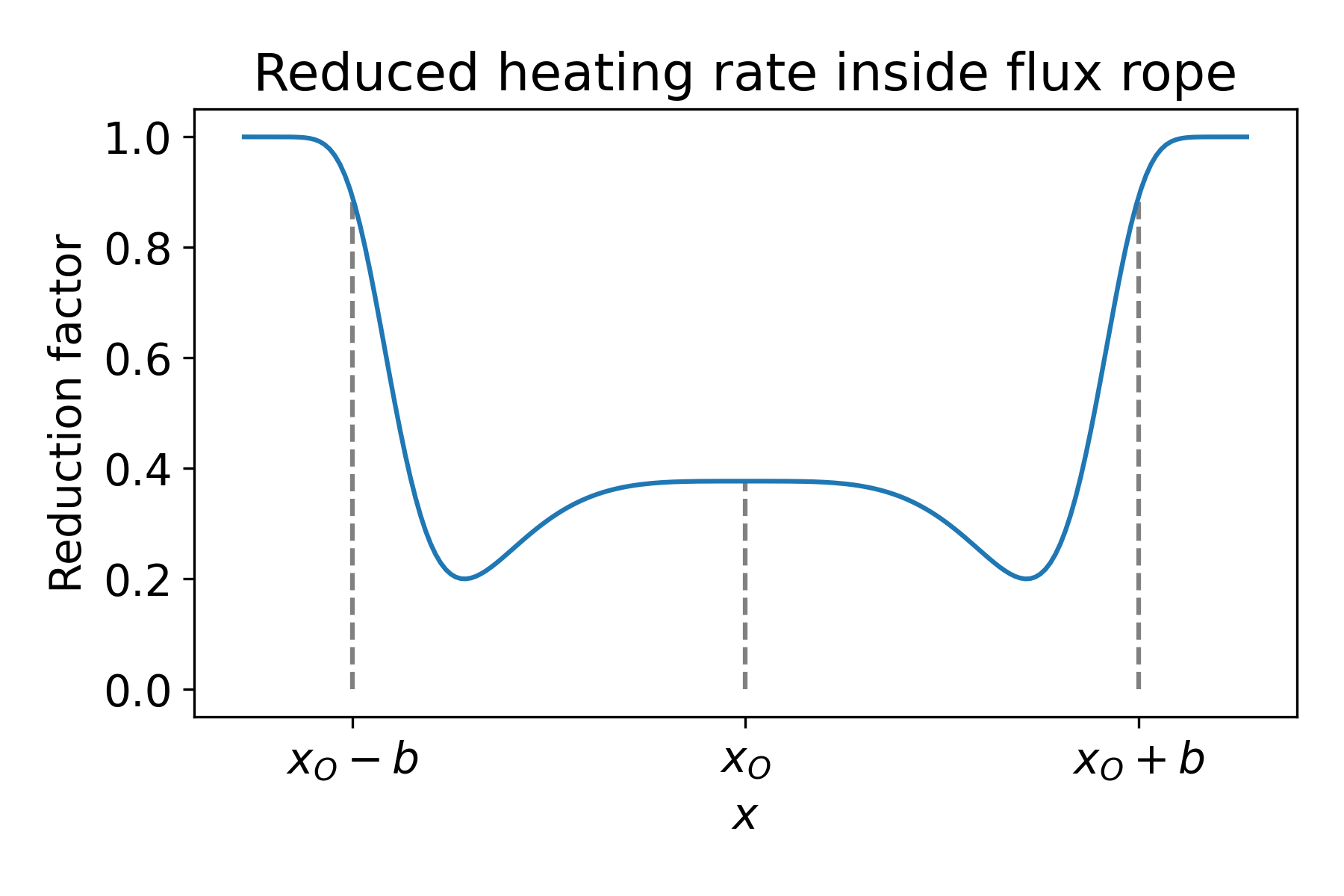}}
	\caption{Cross-section $f(x,y_O)$ of $f$ along the horizontal flux rope axis. Heating is reduced most at the flux rope edges since the field lines are longer. Outside the flux rope, there is no reduction as $f=1$.}
	\label{fig:reduced_heating_f}
\end{figure}

\begin{figure*}
	\centering
	\resizebox{\hsize}{!}{\includegraphics[trim={0 30 0 40 },clip]{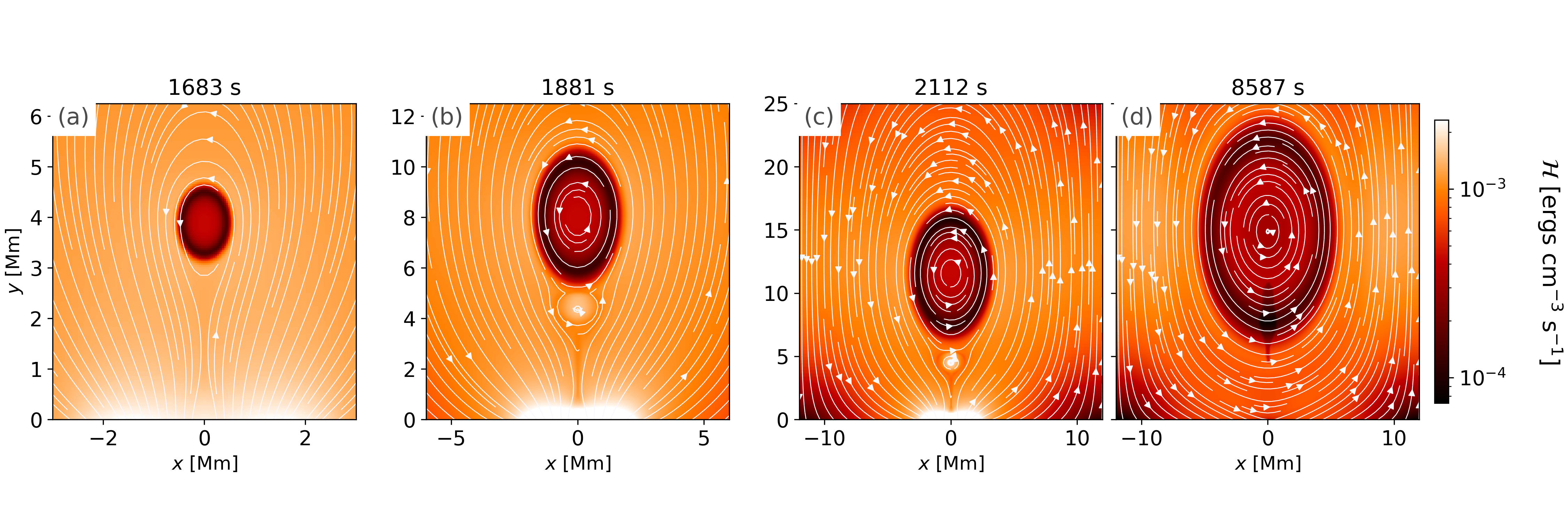}}
	\caption{Demonstration of parameterised heating combined with the reduced background heating  (\texttt{RM1}) inside the flux rope, which is tracked during the simulation. Panels a and b feature zoom-ins of the domain.}
	\label{fig:reduced_heating}
\end{figure*}

The flux rope cross-section at time $t$ is modelled as an ellipse centred at $(x_O(t),y_O(t))$, coinciding with the central flux rope axis, which appears as a magnetic null point (O-point) in the 2D projection. The ellipse's major ($a(t)$) and minor ($b(t)$) axes are given as the distance from the centre to the vertical and horizontal edges of the flux rope, respectively. After these parameters have been determined in the current timestep (see Appendix~\ref{sec:flux_rope_tracking}), the volumetric background heating rate $\mathcal{H}_b$ is transformed following,
%\begin{equation} \label{eq:reduced_heating}
%    \begin{split}
%    & \mathcal{H} = f(\mathbf{u})\times \mathcal{H}_b, \\
%    & f(\mathbf{u}) =  1- \delta e^{-|\mathbf{u}|^4+|\mathbf{u}|^2} /1.284, 
%    \end{split}
%\end{equation}
\begin{equation} \label{eq:reduced_heating}
   \mathcal{H} = f(\mathbf{u})\times \mathcal{H}_b, \quad f(\mathbf{u}) =  1- \delta e^{-|\mathbf{u}|^8+|\mathbf{u}|^4} /1.284, 
\end{equation}
where,
\begin{equation} \label{eq:reduced_coordinate_transform}
	\mathbf{u} = \left( \frac{x - x_O}{0.85b}, \frac{y - y_O}{0.85a} \right),
\end{equation}
is a coordinate transformation centred on the flux rope axis and scaled according to its dimensions. The reduction function $f$ is shown in Fig.~\ref{fig:reduced_heating_f} and takes values between $[1 - \delta, 1]$ since the exponential is normalised by its maximum of $1.284$. To ensure that $f$ does not influence the regions external to the flux rope, we scale $a$ and $b$ by $0.85$ in the coordinate transformation. The above form for the reduction function was chosen so as to reflect the structure of a theoretical twisted flux rope, that is, the field lines at the centre of the flux rope are less twisted and hence shorter than those field lines that form the `boundary' with the corona \citep[cf.][]{Titov:1999}. For a heating rate that varies exponentially along field lines, this leads to modest $\approx$~60\% reduction along the central axis of the flux rope that increases radially away from the central axis up to the flux rope boundary where $f$ returns smoothly to a value of 1 (see Fig.~\ref{fig:reduced_heating_f}). Figure~\ref{fig:reduced_heating} features several snapshots of the reduced heating model \texttt{RM1} applied during runtime, where the flux rope appears as the region of closed poloidal field lines. The volumetric heating rate is seen to be of the order of $10^{-4} - 10^{-3}$ ergs~cm$^{-3}$~s$^{-1}$, which corresponds to values typically taken in literature \citep[e.g.][]{antiochos1991model,dahlburg1998prominence}. At the PIL however, the heating rate greatly increases due to the increased magnetic pressure from bringing the loop footpoints together during the shearing and converging period.

\begin{table*}
	\centering
	\begin{tabular}{lcl|cc|cc}
		\hline\hline
		Heating model & & & Shearing & Condensation & \thead{$n_\textrm{max}$ \\ ($\times 10^{11}$ cm$^{-3}$)} & \thead{$T_\textrm{min}$ \\ (K)} \\ \hline
		\textbf{Exponential} & \texttt{E} & $\rho_0^2 \Lambda(T_0)$ & $+$ & Y & $3.84$ & $7529$ \\
		\textbf{Mixed} & \texttt{M} & $c B^\alpha \rho^\beta$ & & \\
		\phantom{---}Magnetic & \texttt{M0} & $c B^2$ & $-$ & Y & & \\ 
		\phantom{---}Quasi-magnetic & \texttt{M1} & $c B^{2.1}\rho^{-0.1}$ & $+$ & N & &  \\ 
		 			   & & & $-$ & Y & $2.50$ & $8503$  \\ 
		 			   & \texttt{M2} & $c B^{2.2}\rho^{-0.2}$ & $+$ & N & &  \\ 
		 			   & & & $-$ & Y & &  \\ 
		\phantom{---}\cite{Mok2008} & \texttt{M3} & $c B^{2.5}\rho^{-0.5}$ & $-$ & Y & &  \\[0.7ex] 
		 \textbf{Reduced exponential} & \texttt{RE} & $f(\mathbf{u}) \times \rho_0^2 \Lambda(T_0)$ & $+$ & Y & $3.08$
		  & $9196$ \\[0.7ex]
		 \textbf{Reduced mixed} & \texttt{RM1} & $f(\mathbf{u}) \times cB^{2.1} \rho^{-0.1}$ & $+$ & Y & $6.37$ & $8551$ \\[0.7ex] \hline
	\end{tabular}
	\caption{Adopted combinations of heating models and shearing motions with their outcomes. Each model has an abbreviation for further reference. Shearing and anti-shearing motions are denoted by $+$ and $-$, respectively. For those simulations that were advanced far into the prominence evolution, physical values are considered at $t=6741$~s. Both maximum number density and minimum temperature are assumed to be reached in the prominence core.}
	\label{tab:heating_comparison}
\end{table*}

The maximal reduction in heating rate inside the flux rope is controlled by the parameter $\delta$, which is loosely connected to the 3D length of the flux rope. The relation between $\delta$ and $L$ can be approximated assuming that the heating rate $\mathcal{H}$ falls exponentially along field lines: let $s$ be the arc-length coordinate along the central axis of the flux rope, which has length $L$, where our simulation domain consists of a vertical cross-section at $s_0 = 0$. Then the loop footpoints are at $s = \pm \frac{L}{2}$. Assuming the decay length of $\mathcal{H}$ equals the pressure and magnetic scale height $H_0 \approx 50$~Mm, we can derive $L$ from $\delta$:
% \begin{equation} \label{eq:reduced_heating_length}
% 	\begin{split}
% 	& \left( 1 - \frac{\delta}{1.284} \right) \mathcal{Q}\left(\pm \frac{L}{2}\right) \\  
% 	& \qquad = \mathcal{Q}(s_0) = \mathcal{Q}\left(\pm \frac{L}{2}\right) \exp\left(-\frac{|s_0-\frac{L}{2}|}{a}\right)  \\
% 	\Leftrightarrow \qquad & L = - 2a \log(1 - \delta/1.284).
% 	\end{split}
% \end{equation}
\begin{equation} \label{eq:reduced_heating_length}
    \begin{split}
	& \left( 1 - \frac{\delta}{1.284} \right)	= \mathcal{H}(s_0) / \mathcal{H}\left(\pm \frac{L}{2}\right) = \exp\left(-\frac{|s_0\pm\frac{L}{2}|}{H_0}\right)  \\
	& \Leftrightarrow \qquad L = - 2 H_0 \log(1 - \delta/1.284).
	\end{split}
\end{equation}
For the reduced exponential heating model, \texttt{RE} ($\delta=0.8$), this amounts to a flux rope length of $L = 96.6$ Mm, while for the reduced quasi-magnetic heating model, \texttt{RM1} ($\delta=0.9$), we have $L = 119.5$ Mm. These numbers fit well within the observed range of prominence lengths summarised by \cite{Parenti2014}. 
% \nb{Didn't find observations of lengths in Vial et al. 2015.} \jen{The \href{https://ui.adsabs.harvard.edu/abs/2015ASSL..415.....V/abstract}{book}?} \nb{Yes, they mention 100Mm, but don't give any references or range of lengths. Unless I really misread.}.

% Results section

\section{Results} \label{sec:results}

In our simulations, a flux rope forms through multiple reconnection events, creating a 2D structure of poloidal nested flux surfaces, alternately hotter and cooler as a result of mergers of multiple subordinate flux ropes -- a detailed description can be found in \cite{jack2020}. With the flux rope fully formed after $2800$~s, the energy injection into the plasma as a consequence of the formation process ceases. From then on, the thermodynamic properties of the dense material suspended by the concave upwards field are thus governed primarily by the ratio between the heating and cooling terms of the energy equation. Under the conditions that the local cooling term dominates over the combined influence of the heating or conduction terms, the associated plasma may become unstable to the thermal instability. However, and as we shall henceforth show, when, how, and whether this happens at all depends directly on the adopted heating model in combination with the (anti-)shearing motions involved in the initial formation of the flux rope.

\subsection{Condensation phase} \label{sec:condensation_phase}

Table~\ref{tab:heating_comparison} summarises the different combinations of heating models and footpoint motions under consideration and whether thermal instability occurs, in analogy with \citep[Table 1]{KY2015}. In Sects.~\ref{sec:prominence_exponential}~--~\ref{sec:prominence_reduced_mixed}, we look at how the flux rope material reacts to these different prescriptions.

\subsubsection{Exponential heating model} \label{sec:prominence_exponential}

Model \texttt{E} contains a steady background heating rate that decreases exponentially with height. We consider a flux rope formed through positive shearing motions, as in \cite{jack2020}, and find similar overall evolution: a dense cloud suspended by the magnetic field collapses following the condensation process described in \cite{ClaesN2019Tsom}. That is, radiative losses increase, resulting in a decrease of temperature and increase of density, which again enhances the radiative losses and leads to a runaway evolution; matter is accreted along each flux surface, producing a condensation perpendicular to the magnetic field, typical of the thermal instability in 2D \citep{ClaesN2019Tsom,joris2021}. After the two main condensations form at around $y = 3$~Mm and $4.5$~Mm, some smaller condensations appear at various locations along the outer flux surfaces of the flux rope. These smaller blobs then fall down along the magnetic field under the effect of gravity, reaching speeds up to $190$~km~s$^{-1}$ and eventually join the larger condensation. Imbalances in the velocities of inflowing material deform the smaller condensations as they move downwards. The momentum of the plasma gained through falling results in sloshing motions of the prominence body, with distinct oscillation periods along each flux surface. Towards the end of the simulation, the material slows down and the condensed material reaches relative equilibrium, during which the prominence sinks as a whole due to enhanced ($\eta = 0.002$) mass slippage over the field lines \citep{low2012a,jack2020}.

\subsubsection{Mixed heating models} \label{sec:prominence_mixed}

\begin{figure*}
	\centering
	% \resizebox{\hsize}{!}{\includegraphics{Figures/mix_as_qm_evolution.png}}
	\resizebox{\hsize}{!}{\includegraphics[draft=false]{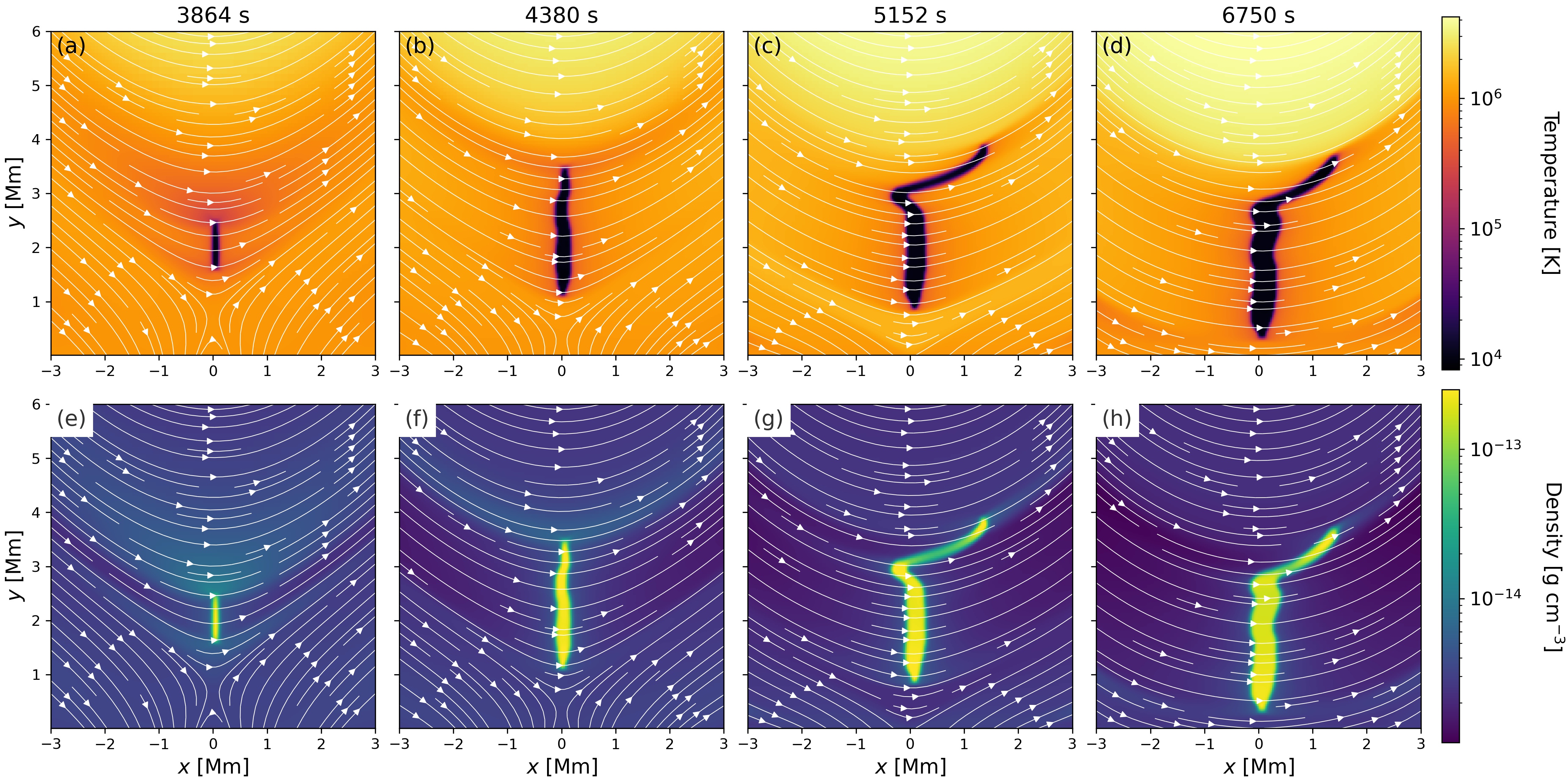}}
	\caption{Density and temperature evolution for the simulation performed with quasi-magnetic heating (\texttt{M1}) and anti-shearing motions. One monolithic condensation forms, growing and descending at a later stage in its evolution. A movie of this figure, where the coronal rain blobs outside of the flux rope are also featured, is available with the online version of this manuscript.}
	\label{fig:mix_evolution}
\end{figure*}

The similar study of \cite{KY2015} previously concluded that in situ prominence formation is inhibited when positive shearing motions are coupled with a heating model $\mathcal{H}\sim B$, while coupling the motions with a model $\mathcal{H}\sim\rho$ does produce condensations. In agreement with their results, we find no condensations in our positive shearing simulations with mixed $-0.2\leq~\beta~\leq~0$ models \texttt{M0}, \texttt{M1}, \texttt{M2}, but rather hot material at a temperature of about $6.3$~MK. Since the temperature of the plasma within the flux rope constructed using the positive shearing \texttt{M2} model was so high, we did not test a positive shearing \texttt{M3} configuration with $\alpha = 2.5$. Forming the flux rope through anti-shearing motions, however, does produce condensations. These have similar characteristics for all models, yet are quite distinctive from those obtained with the exponential heating model. In particular, the results for the quasi-magnetic heating model $(\alpha,\beta)=(2.1,-0.1)$ closely match those for the magnetic heating model $(\alpha,\beta)=(2,0)$. For \texttt{M0}\,--\,\texttt{M3}, thermal instability initiates at around the same time ($t=3250$~s). Figure \ref{fig:mix_evolution} shows the density and temperature evolution for the quasi-magnetic \texttt{M1} model with anti-shearing motions. 
Here, one large monolithic condensation forms at the bottom of the flux rope. Post-formation, the prominence is seen to expand and descend (Panels c and d of Fig.~\ref{fig:mix_evolution}), while its top moves due to asymmetric inflows. Unlike in the other simulations, parts of the bottom atmosphere become thermally unstable due to the background heating decrease by the anti-shearing motions. As the resulting coronal rain forms outside of the flux rope, it falls down along the open field lines. However, the bottom boundary in our setup does not allow the material to leave the coronal domain as a chromosphere has not been included, and hence the condensations remain unphysically suspended above the bottom boundary. As the prominence body grows, its lower end makes contact with the flux surface on which these condensations reside. To allow for an analysis of the prominence characteristics, we applied an additional consistency check (for $\nabla\cdot\mathbf{B}=0$) to the bottom boundary, which slightly increases the temperature inside the subordinate flux ropes due to increased dissipation of localised currents. This subsequently disconnects the prominence material from these ‘coronal rain’ blobs, and enables an analysis of the prominence material up to up to t=6741 s. After this time, the blobs do connect to the flux surface of the main prominence body and we end the analysis.

\subsubsection{Reduced exponential heating model} \label{sec:prominence_reduced_exponential}

Model \texttt{RE} combines the exponential heating model with the reduction function $f$ that dynamically modifies the heating rate inside the flux rope, as explained in Sect.~\ref{sec:heating_reduced}. One can thus expect that the simulations will be identical up until the flux rope formation is initiated ($t=1300$~s). We therefore initialise the \texttt{RE} simulation with the one for model \texttt{E} at that time. Since we choose $\delta = 0.8$, the heating rate is reduced in the core of the flux rope to approximately 40\% of its original value, and to 20\% near the edges. 

\begin{figure*}
	\centering
		\resizebox{\hsize}{!}{\includegraphics[draft=false]{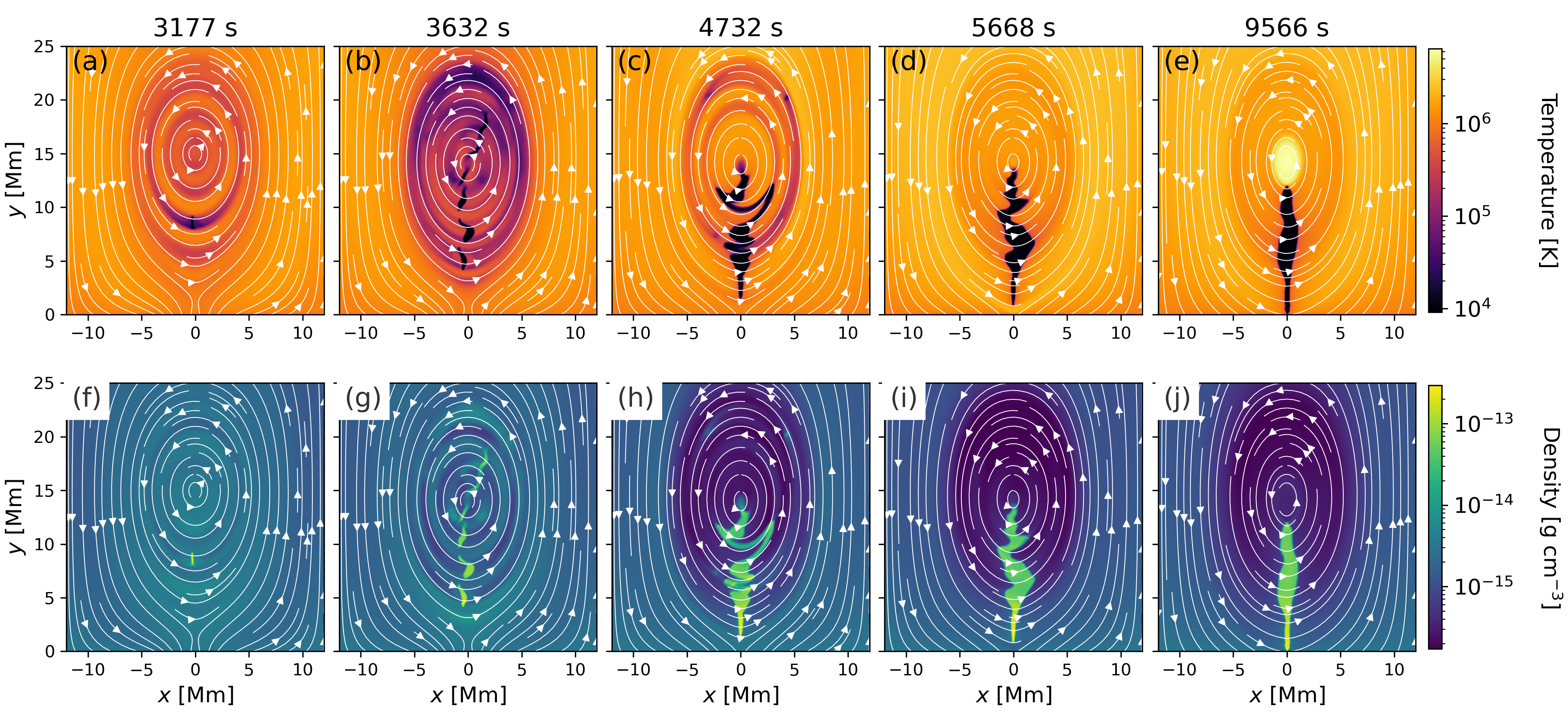}}
	\caption{Density and temperature for flux rope formed by shearing motions with reduced exponential heating model. All over the flux rope, condensations form and fall down along the magnetic field, collecting in one large prominence body. A movie of this figure is available with the online version of this manuscript.}
	\label{fig:exp_red_evolution}
\end{figure*}

Figure \ref{fig:exp_red_evolution} shows the first large condensation to form around $3000$~s at a height of around $8$~Mm. Generally speaking, condensations occur in every other flux surface, as a result of the temperature alternating between hotter-cooler due to the preceding mergers of several secondary flux ropes. Some condensations consist of only one blob, some consist of a blob on either side of the central vertical axis which fall down and collide. After the collision, a shock-like feature is seen to propagate away from the impact location which resembles the shocks seen at the collapse of condensations within previous simulations of the thermal instability \citep[cf.][]{ClaesN2020TiFa}.
% the interfaces of which appear to move with velocities $11 - 20$~km/s. 
Thermal instability remains relevant up to $t=4700$~s, after which no additional condensations form as most of the dense matter bound by the flux rope has already cooled down.

% Schlieren plot of 'shocks' was inconclusive, but in any case they appear for both symmetric condensations colliding and monolithic condensations. Might not exactly be shocks, but, e.g., mass `overshooting'?

Most condensations fall down rapidly along field lines under the effect of gravity, overshoot the concave-up portion of the domain due to their finite kinetic energy, before subsequently performing damped oscillations about $x=0$ with different periods along each field line. Some smaller blobs move upwards as a result of upflows and changing pressure conditions as a result of material evolving elsewhere within the flux rope. This is, however, a consequence of the 2.5D description of the flux rope, where the poloidal field line of a flux surface connects plasma elements that would not necessarily be connected in a 3D description of the flux rope. Strikingly, both of these condensations eventually evaporate while falling down at around $5000$~s before reaching the main prominence body. After these condensations disappear, much of the material inside the flux rope has collected in the extended prominence, whereas the density of the hot material in the flux rope drops to about $20\%$ of the ambient coronal density. As highlighted well in the online movie that accompanies Figure~\ref{fig:exp_red_evolution}, the large prominence appears to have a significant effect on the hosting flux rope, clearly deforming the field line topology despite the strong $10$~G field. As in the simulations with the mixed heating model, the bottom of the prominence sinks and compresses field lines, leading to an increased perpendicular current density. The outer flux rope shell is heated upon compression by the material at the bottom of the domain, as is evident from Panel~d of Fig.~\ref{fig:exp_red_evolution}. At the same time, the temperature at the flux rope centre greatly increases through resistive heating and mass slippage of the prominence top.

%Unlike in our other simulations, the flux rope oscillations have now a horizontal component, due to the angular momentum of the large amount of material collected in the prominence.
%The whole prominence falls at a rate of $0.11$ km/s.

\subsubsection{Reduced quasi-magnetic heating model} \label{sec:prominence_reduced_mixed}

\begin{figure*}
	\centering
	\resizebox{\hsize}{!}{\includegraphics[draft=false]{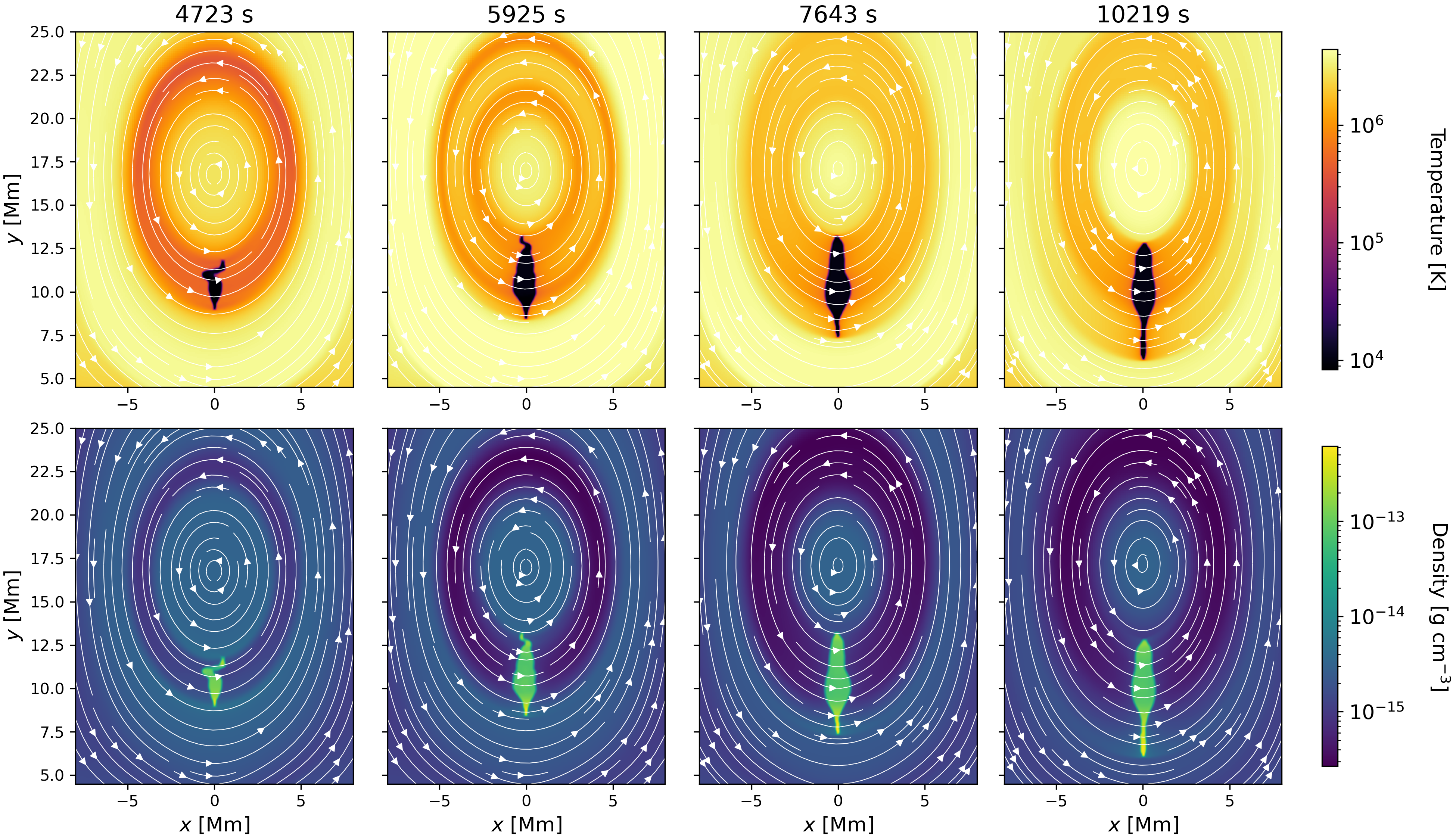}}
	\caption{Density and temperature for flux rope formed by shearing motions with reduced quasi-magnetic heating model. One monolithic condensation forms at the bottom of the flux rope, visibly containing more mass than for the unreduced mixed heating model in Fig.~\ref{fig:mix_evolution}. A movie of this figure is available with the online version of this manuscript.}
	\label{fig:mix_red_evolution}
\end{figure*}

Finally, we perform a simulation with the reduced quasi-magnetic heating model (\texttt{RM1}) and a flux rope formed through shearing motions. In Sect.~\ref{sec:prominence_mixed}, it was shown that models \texttt{M} combined with shearing motions produce too hot a flux rope for thermal instability to occur, analogous to \cite{KY2015}. However, observational studies have long asserted that shearing motions are ubiquitous to PILs and hence transfer complimentary shear to any associated magnetic arcades that straddle this divide \citep[e.g.][]{Athay:1986}. Remarkably, through the dynamic cross-sectional heating reduction, this contradiction is overcome, although outside of the flux rope the heating model does still lead to high temperatures. As can be seen in Fig.~\ref{fig:reduced_heating}, the heating rate at the vertical edges at $|x|=12$ Mm increases up to order $10^{-3}$ ergs s$^{-1}$ cm$^{-3}$ due to low density and high magnetic pressure, leading to downflows that further decrease the local density. One filamentary condensation appears at the bottom of the flux rope (Fig. \ref{fig:mix_red_evolution}) around $3850$~s.

The condensation closely resembles those obtained with models \texttt{M} and anti-shearing motions, but is larger. Again, we observe shock-like features emanating from the initial cloud out along the magnetic field when it collapses. Even though the heating rate is reduced, thermal instability remains absent from the central region of the flux rope, which consequently only contains hot material. The prominence is seen to grow from its lower end, probably as a consequence of the high-density region at the bottom, which reduces the local heating rate in model \texttt{M1}, in combination with the mass-slippage mechanism previously remarked upon within these models \citep{low2012a,jack2020}. 
% \nbn{}{Towards the end of the simulation, a zig-zag like deformation appears at the top of the thinner prominence part. A local increase in perpendicular ram pressure $\rho v^2$ suggests that this might be the thin shell instability at play, analogous to the breaking up of filamentary structures reported by \cite{joris2021}.}

\subsection{Comparison of the heating models}

% Maybe move part of this to discussion.
Table~\ref{tab:heating_comparison} summarises the outcomes for the combinations of heating models and footpoint motions considered, analogous to Table~$1$ in \cite{KY2015}. We find those heating models based on local parameters to require anti-shearing motions in order to produce conditions suitable for prominence formation. Applying the ad hoc, masked heating reduction inside the flux rope re-enables the possibility of condensation formation with shearing motions.

\begin{figure}
	\centering
	\resizebox{\hsize}{!}{\includegraphics[draft=false]{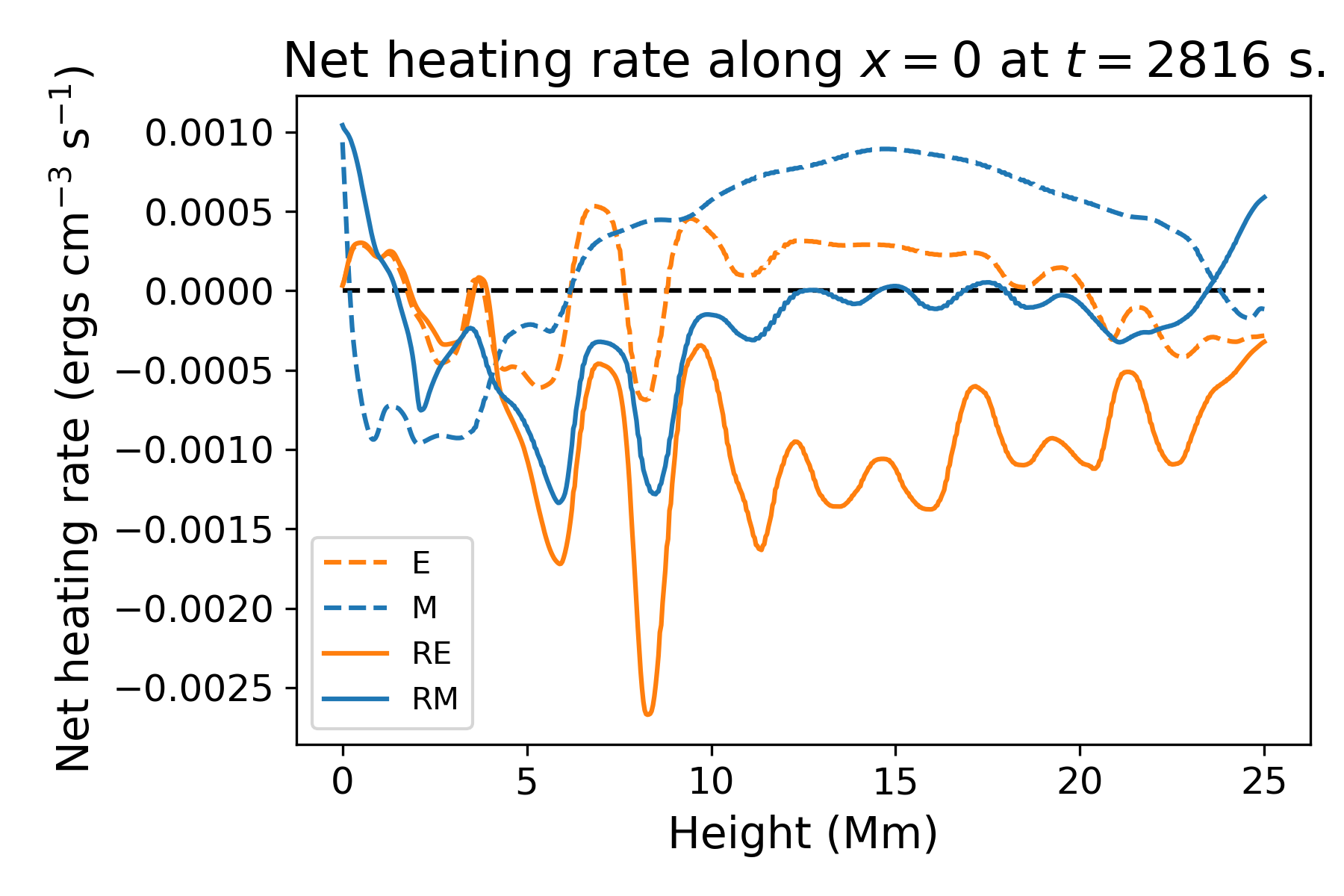}}
	\caption{Cut along the $y$-axis showing the net heating rate, $-\rho\mathcal{L}$, when the driving motions end. Regions where the energy transfer rate is negative are more prone to thermal instability.}
	\label{fig:net_heating}
\end{figure}

The locations of condensation formation vary between the heating prescriptions but remain governed by the imbalance between heating and cooling and the stabilising effect of thermal conduction. Figure~\ref{fig:net_heating} features the net heating rate along a vertical cut over the $y$-axis, immediately after the flux rope has fully formed. The regions of negative net heating rate at this instant indeed correspond well to those flux surfaces that ultimately experience the in situ condensation shown in the first panels of Figs.~\ref{fig:mix_evolution}, \ref{fig:exp_red_evolution}, and \ref{fig:mix_red_evolution}. In particular, model \texttt{RE} is seen to lead to a net cooling rate over the entire flux rope, which enables most of the levitated material to cool and rain down.

\begin{figure}
	\centering
%	\begin{subfigure}{0.49\textwidth}
%		\includegraphics[width=\textwidth]{Figures_driven/Driven_Tmax.png}	
%	\end{subfigure}
%	\begin{subfigure}{0.49\textwidth}
%		\includegraphics[width=\textwidth]{Figures_reduced/reduced_maxT.png}	
%	\end{subfigure}
	\resizebox{\hsize}{!}{\includegraphics[draft=false]{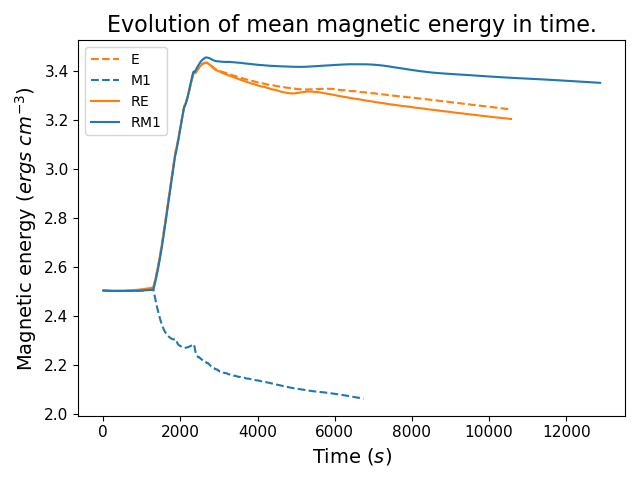}}
	\caption{Mean magnetic energy density evolution over the domain for models \texttt{E}, \texttt{M1}, \texttt{RE} and \texttt{RM1}. Shearing motions increase the magnetic energy, anti-shearing motions lead to it decreasing.}
	\label{fig:all_magnetic}
\end{figure}

The evolution of the mean magnetic energy density over the domain for the four selected simulations that produced condensations, detailed in Fig.~\ref{fig:all_magnetic}, shows clearly the distinction between shearing and anti-shearing footpoint motions: the former store energy in the magnetic field while the latter decrease the magnetic shear and hence also the value of $B_z$. In the initial relaxation phase the magnetic energy evolution is comparable for all simulations. The abrupt change at $1300$~s is due to the activation of the driven boundary motions, bringing the footpoints of the initial arcade together and hence driving an energetic evolution of the system. On a related note, there appears to be a relationship between the direction of shearing motions and the number of additional subordinate flux ropes appearing with the current value of $\eta$: shearing motions consistently lead to the formation of four large subordinate flux ropes that merge with the main flux rope, while anti-shearing motions produce only two. 

An analysis of the maximum temperature over the simulation domain reveals the influence of the heating models on the flux rope and the ambient atmosphere. The mixed models can be ranked according to the power $\alpha$ for the magnetic field strength: a higher $\alpha$ produces a hotter atmosphere and flux rope regardless of shearing direction. Ordered by increasing maximum temperature, we have \texttt{M0}--\texttt{M1}--\texttt{M2}--\texttt{M3}. When the flux rope has fully formed after $2800$~s, the maximal temperatures within the mixed models are of the order of $8$~MK. For the models with reduced heating, the temperatures are then much lower at around $4 - 5$~MK. This is clearly the influence of the diminished heating rate inside the flux rope. For model \texttt{RE}, the maximum temperature also increases at a slower pace due to the lower heating rate inside the flux rope: the centre of the flux rope therefore heats mostly due to resistive Ohmic dissipation (Joule heating). Another effect is that the maximum temperature of model \texttt{RE} levels out soon after the shearing motions end, while it continues to increase for the other heating models. However, at around $5000$~s, resistive heating in the flux rope centre becomes more significant and increases the temperature out of equilibrium for model \texttt{RE} as well.

%Starting from approximately $30$ minutes, the maximum temperature evolution shows several spikes: two for the mixed models and three for the exponential heating model and reduced models respectively. These are a result of the reconnection events associated with the formation of (secondary) flux ropes described in Sect. \ref{sec:flux_rope_formation}. This confirms that anti-shearing footpoint motions (here for the mixed models) lead to the formation of only two major subordinate flux ropes while shearing motions produce three major subordinate flux ropes. Minor reconnection events were also reported, which show up as small peaks on the maximum temperature evolution (not visible on Fig. \ref{fig:all_magnetic}). 

%   For the quasi-magnetic heating model, boundary effects lead to an increase in temperature up to $10$ MK in the outer regions of the flux rope, but after $2$ hours they slowly disappear. At around $2.7$ hours, the maximum temperature traces the temperature of the flux rope centre as evident from the discontinuous transition in the slope of the temperature evolution. It reaches temperatures of up to $5.1$ MK, comparably lower than for the quasi-magnetic model without heating reduction. 

\begin{figure}
	\centering
	\resizebox{\hsize}{!}{\includegraphics[draft=false]{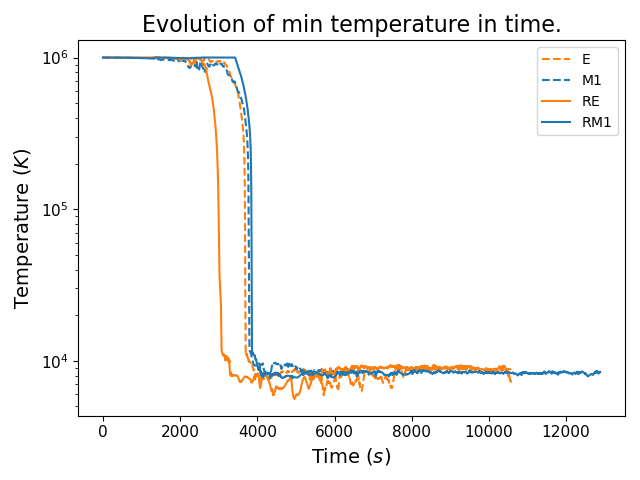}}
	\caption{Evolution of the minimum temperature for models \texttt{E}, \texttt{M1}, \texttt{RE} and \texttt{RM1}. The minimum temperature decreases non-exponentially to less than $10^4$ K, with apparently higher growth rates for the reduced models.}
	\label{fig:comparison_models_Tmin}
\end{figure}

We present the minimum temperature evolution for models \texttt{E}, \texttt{M1}, \texttt{RE} and \texttt{RM1} within Fig.~\ref{fig:comparison_models_Tmin}. This temperature is exclusively reached inside the condensations and prominence body, which we limited to $|x| \leq 3$~Mm for the run with model \texttt{M1} as to exclude the coronal rain from the analysis. The late phase, post-condensation temperatures are found in the range $6000 - 10\,000$~K, which are values well above the artificial minimum of $10^3$~K adopted in our simulations. For both the original and the reduced heating models, the minimum temperature decreases non-exponentially, although a satisfactory exponential fit can be obtained for models \texttt{E} and \texttt{M1} nonetheless. This is not unexpected, as the reduced heating models impose conditions out of thermal equilibrium, which could explain the sharp decrease in temperature as a non-linear perturbation. It is hence non-trivial to map the initial growth rate of thermal instability to the evolution into the non-linear regime, as was previously possible in the thermal instability simulations of \cite{ClaesN2019Tsom}. Nevertheless, we find approximate growth rates $\omega_i = 0.002$~s$^{-1}$ (model \texttt{M1}), $\omega_i = 0.004$~s$^{-1}$ (model \texttt{E}) and $\omega_i \sim 0.007$ (\texttt{R}-models), which agrees well with an estimate from linear theory based on the local parameters, and neglecting thermal conduction and resistivity, of $\omega_i \sim 0.001$~s$^{-1}$. 

\begin{figure*}
	\centering
%	\begin{subfigure}{0.49\textwidth}
%		\includegraphics[width=\textwidth]{Figures_reduced/reduced_minT.png}	
%	\end{subfigure}
%	\begin{subfigure}{0.49\textwidth}
%		\includegraphics[width=\textwidth]{Figures_driven/Driven_Tmin.png}	
%	\end{subfigure}
    \begin{subfigure}{0.75\textwidth}
	\resizebox{\hsize}{!}{\includegraphics[draft=false]{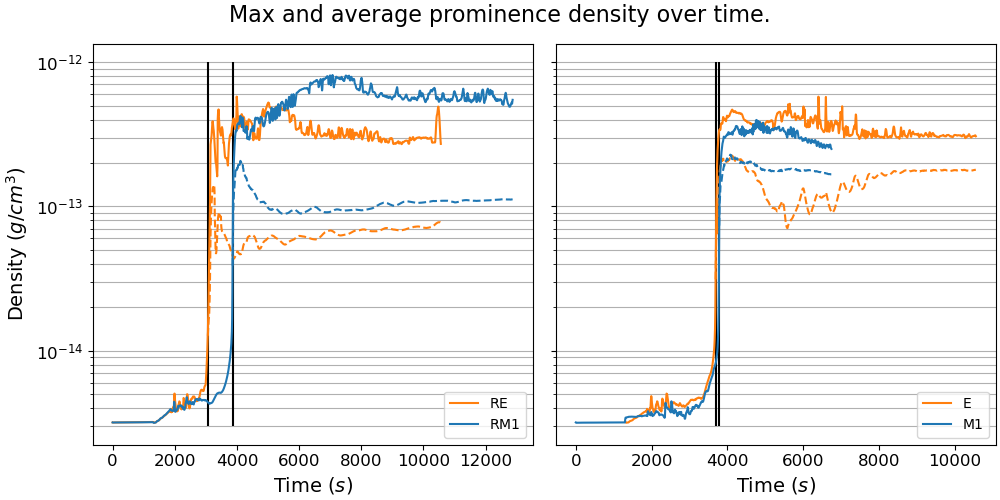}}
	\end{subfigure}
	\caption{Evolution of maximum (solid) and average (dashed) density for models \texttt{E}, \texttt{M1}, \texttt{RE} and \texttt{RM1}. Black line indicates the appearance of the first condensation.}
	\label{fig:comparison_models_densities}
\end{figure*}

The evolution of the average prominence, and maximum domain densities for models \texttt{E}, \texttt{M1}, \texttt{RE}, and \texttt{RM1} is presented in Fig.~\ref{fig:comparison_models_densities}, wherein the occurrence of the first condensation is marked with a vertical black line. We see that the reduced heating model \texttt{RE} leads to the development of condensations at an earlier time than model \texttt{E}. For models \texttt{M1} and \texttt{RM1}, this behaviour is flipped as a result of the accompanying shearing motions in the second case, which lead to less favourable conditions to in situ condensation since the field strength and hence heating rate increases.

%\jer{Next, we look at the maximum density over the domain, and average prominence density, featured on Fig.~\ref{fig:comparison_models_densities} for the four heating models considered above.} 
To calculate the average prominence density, we define the prominence material as having: $T < 25\,000$~K and $\rho > 1.171 \times 10^{-14}$~g/cm$^3$ ($=50$ code units), and in the case of model \texttt{M1} limit the analysis to $|x| \leq 3$~Mm. In all cases, the maximum density is reached at the bottom of the prominence, as a result of the larger-area flux surface encapsulating more material. The small peak towards the end for model \texttt{RE} is, however, reached in the prominence top when it compresses due to field slippage. We find both the maximum density of condensations produced by model \texttt{RM} to be consistently higher than those obtained with the other models. This is most likely related to the nature of the mixed model, which reduces heating in regions of increased density and hence the material is more free to cool and contract in turn. A similar phenomenon may occur with model \texttt{M} if the simulation were extended without interference by the coronal rain.
% \RK{We find the use of the ad hoc heating reduction to have a minimal effect on the maximum densities attained for the models considered here, in each case reaching the same order of magnitude.}{} \jen{I think this is still important to note, since it agrees with K\&Y2017, but perhaps we already repeat it somewhere else?} 
In terms of average prominence density, models \texttt{E} and \texttt{M} are found to outclass those for the reduced models by a factor two. We find this to be a consequence of the larger condensations created by the reduced heating models, where the additional volume is provided by material of lower density.
% situated above the compressed material at the lower end of the prominence. 
Hence, the distribution of prominence densities with these modified models is concentrated around lower values (see also Sect. \ref{sec:towards_realism}).

\begin{figure}
	\centering
	\begin{subfigure}{\hsize}
		\resizebox{\hsize}{!}{\includegraphics{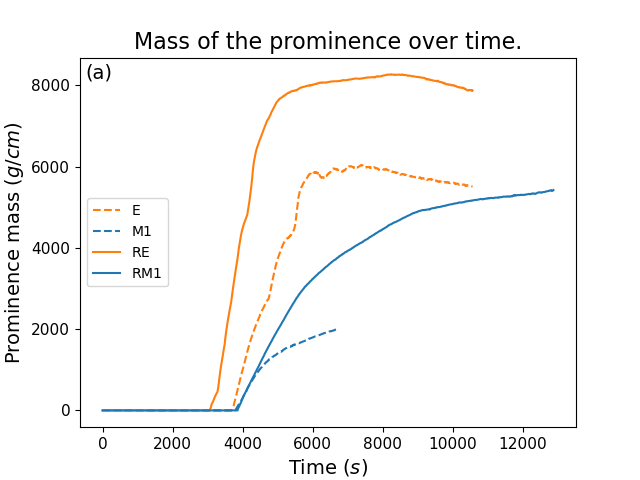}}
	\end{subfigure}
	\begin{subfigure}{\hsize}
		\resizebox{\hsize}{!}{\includegraphics{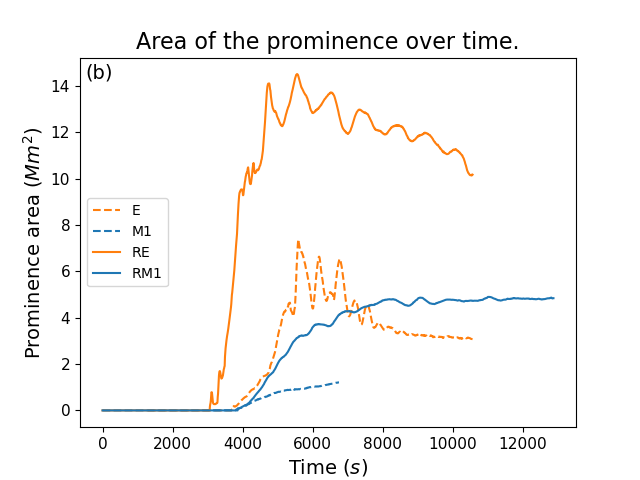}}
	\end{subfigure}
	\caption{Evolution of the total prominence mass per unit length and its 2D surface area, for the exponential and quasi-magnetic heating models (dashed) and their reduced counterparts (solid). Both in terms of area and mass, the reduced heating models lead to larger prominences.}
	\label{fig:comparison_models_area_mass}
\end{figure}

We present the evolution of both prominence mass and area for models \texttt{E}, \texttt{M1}, \texttt{RE}, and \texttt{RM1} in Fig.~\ref{fig:comparison_models_area_mass}. We note immediately that the two reduced heating models produce larger and more massive prominences compared to the `original' models, with a mass increase of about $30\%$ for model \texttt{E} and $160\%$ for model \texttt{M1}, and increase in area by a factor three. A similar observation applies to models (\texttt{R})\texttt{E} compared to models (\texttt{R})\texttt{M}: the mixed models produce significantly less massive condensations, with even the reduced heating model \texttt{RM1} being barely able to outpace model \texttt{E}. The resulting condensations are however comparable in terms of area. Mass increase with model \texttt{RE} stabilises around $6000$~s when the last condensations have either rained down or evaporated. For models (\texttt{R})\texttt{E}, the prominence mass stabilises only briefly before decreasing due to evaporation of material within the prominence `monolith', at rates of $-0.2$ and $-0.15$~g~cm$^{-1}$~s$^{-1}$, respectively. In the later stages of evolution, both simulations with the exponential models show a decreasing area as a consequence of the concentration of plasma at the concave upwards portions of the magnetic topology over time. Under these conditions the extent of the plasma along a flux surface is increasingly governed solely by the local pressure-scale height. 

In addition to the size of condensations, the total area evolution also contains information on the oscillations of condensed material in the dipped magnetic field as mentioned above and visible in Figs.~\ref{fig:exp_red_evolution}~and~\ref{fig:mix_red_evolution}. Extracting the period of the largest oscillatory motions from the periodicity in the area evolution, we find dominant periods of $335$~s for model \texttt{E}, $904$~s for model \texttt{RE} and $840$~s for model \texttt{RM1}. We explore this in more detail in Sect.~\ref{sec:additional_results}. The quasi-magnetic heating model \texttt{M1} does not lead to any oscillations since the single condensation forms in-place at the locations of concave-up magnetic topology.

%Although not all simulations with mixed heating models were advanced far past the condensation phase of prominence formation, we can compare temperatures at a common timestamp and find an ordering of the mixed heating models in terms of prominence temperature: if the heating rate depends on a larger power $\alpha$ of the local magnetic field strength, then the resulting condensation will be hotter. 

%\begin{table*}
%	\centering
%	\begin{tabular}{cccc}
%		\hline\hline
%		\textbf{Heating model } & $\rho_\textrm{max}$ ($\times 10^{-13}$ g cm$^{-3}$) & $n_\textrm{max}$ ($\times 10^{11}$ cm$^{-3}$) & $T_\textrm{min}$ (K) \\  \hline
%		\textit{Exponential} & 3.14 & 4.46 & 8904 \\ 
%		\textit{Quasi-magnetic} & 4.89 & 4.87 & 9274 \\ 
%		\textit{Reduced exponential} & 3.08 & 3.81 & 8437 \\ 
%		\textit{Reduced quasi-magnetic} & 6.06 & 5.05 & 9254 \\ \hline
%	\end{tabular}
%	\caption{}
%	\label{tab:prominence_characteristics}
%\end{table*}

\subsection{Phase space evolution}

\begin{figure*}
	\centering
	\includegraphics[width=17 cm, draft=false]{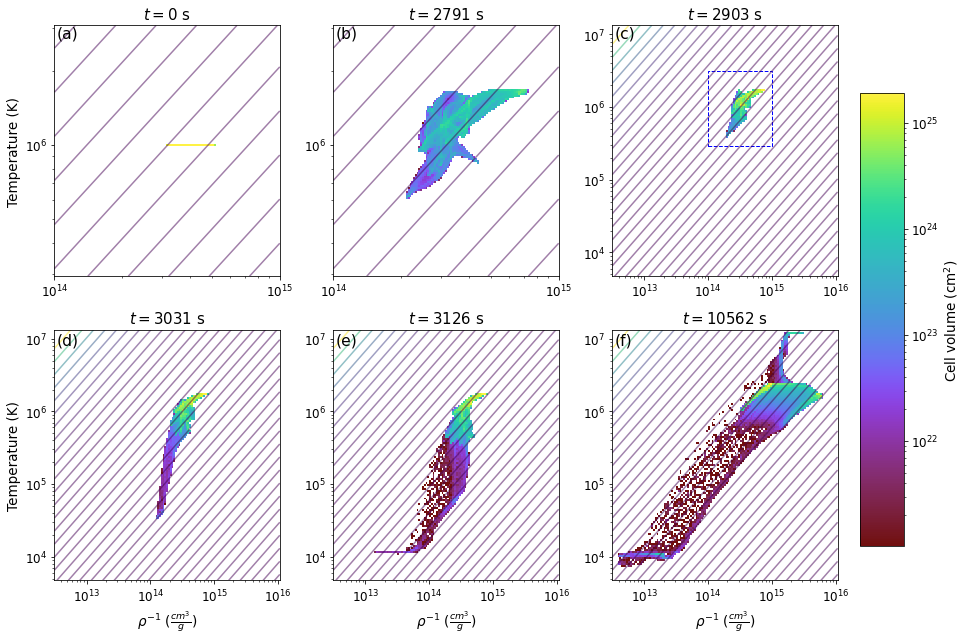}
	\caption{Evolution in $\rho^{-1}-T$ state space for model \texttt{RE}, with pressure isocontours diagonally. Distribution given in total cell volume. A movie of this figure is available with the online version of this manuscript.}
	\label{fig:phase_space}
\end{figure*}

Criteria for the initiation of the thermal instability have been derived under both isochoric and isobaric criteria \citep{parker, field}. To ascertain which of these approximations is more consistent with the evolution found within our prominence condensations, we now look at the $\rho^{-1} - T$ phase space, shown in Fig.~\ref{fig:phase_space}, at several snapshots for the simulation with model \texttt{RE}. The coloured diagonal lines indicate the pressure isocontours governed by the normalised ideal gas law $p = \rho T$. The distribution is coloured by total volume of cells in the simulation domain that correspond to a certain bin in phase space, and hence contains information on the physical extent of regions in phase space. A similar analysis was performed by \cite{waters2019} for unmagnetised astrophysical fluids. In contrast to their analysis, we do not use the concept of an `S-curve' of thermal equilibrium conditions, but instead the qualitative description of the phase evolution of thermal instability.

Initially, the atmosphere is isothermal with a varying density profile due to hydromagnetostatic equilibrium, which translates to a horizontal line in the $\rho^{-1} - T$ diagram (Panel~a). As time progresses, the atmosphere begins to heat and the distribution in phase space takes on complex shapes when the coronal loop footpoints are driven. After some time, the thermal instability is initiated and a branch towards the cold, dense lower left corner of the state space domain splits off (Panels~c and~d). The abrupt cooling of the material seems to follow the pressure isocontours fairly well in the initial stage at $t=2903$~s, but the branch is seen to cross several isocontours soon afterwards and follows a more vertical and hence more isochoric path. In Panel~e, the evolution of the condensation at its coolest point happens under almost isothermal conditions as a consequence of the significant drop in radiative losses at $\approx 10^4$~K. In the final snapshot on Panel~f, the hot, tenuous flux rope and relatively colder, massive condensation appear as two distinct populations in the lower left and upper right corners of the distribution, respectively. In fact, two hot and tenuous populations can be distinguished as two lines of increased volume aligned with the pressure isocontours and hence at approximately uniform pressure. The left, upper population corresponds to the atmosphere surrounding the flux rope, while the right, lower population corresponds to the flux surfaces on which the prominence resides. The latter are tenuous because much of the material has condensed. The hot flux rope centre appears as a vertical extension in the right upper corner of the phase diagram.

% similarly to \cite{waters2019}. 

\section{Discussion}\label{sec:discussion}

We performed 2.5D simulations of prominence formation through the levitation-condensations mechanism, where the basic simulation setup is taken as in \cite{jack2020}. The original model of \cite{KY2015} has been further modified to include a density stratification profile that is fully consistent with the varying gravitational acceleration \eqref{eq:hydrostatic_profiles} over the domain. A value of $B_a = 10$~G was taken for the background magnetic field at the bottom of our simulation domain while the simulation of \cite{jack2020} at the same resolution had a $3$~G background field. This has implications for the magnitude of Ohmic heating at the flux rope centre and hence the amount of material that is able to condense.

\subsection{Influence of heating models on prominence formation}

\subsubsection{Exponential heating models}

The simulation performed with the exponential heating model \eqref{eq:heating_exponential} closely matches the results from \cite{jack2020}. However, in our case the material collects in one main prominence body while it remained disconnected in different flux surfaces in the $10$~G field simulations of \cite{jack2020}. This is most likely a consequence of the lower resolution taken in our simulations, which leads to a slightly lower heating input from reconnection of subordinate flux ropes. Nevertheless, the large central part of the flux rope remains as a dense and hot region insusceptible to thermal instability, as evident from the heating-cooling balance in Fig.~\ref{fig:net_heating}. Indeed, following the formation of the flux rope, the material bound by the flux rope should be isolated from the strongest contributions of the background heating as thermal conduction can no longer transport this energy along field lines. Therefore, we introduced a method to implement an ad hoc modification to the heating profile so as to approximate the influence of field line connectivity to the heating experienced within the 2.5D flux rope core.

This reduced exponential heating model leads to a significantly cooler flux rope than its unmodified counterpart, enabling radiative cooling to easily dominate over background heating. Strikingly, nearly all of the material in the flux rope is found to cool and rain down into the prominence, turning the flux rope into a cavity with densities of the order of $10^{-16}$ g cm$^{-3}$. Although similar results were found before, this is the first time it has been demonstrated with a field strength as high as $10$~G. The flux rope centre is much cooler than in previous simulations without reduced heating and even hosts in situ condensations. In the final stages of the simulation, the flux rope temperature increases to order $10$~MK which is rather high compared to observed cavity temperatures of $1.5 - 2.2$ MK \citep{bak2019thermal}. Such an evolution is compounded by the high resistivity value in our simulations leading to increased Joule heating in the cavity combined with increased mass slippage at the prominence top.

% Towards the end of the simulation, resistive heating again takes over and increases the temperature in the flux rope centre. However, the region of increased temperatures is limited to the flux surfaces that do not contain prominence material and so this does not represent an evaporation process.

\subsubsection{Mixed heating models}

The original study of \cite{KY2015} investigated the effect of background heating $\mathcal{H} \sim B^2$ and $\mathcal{H} \sim \rho$ in a similar setup. They used a background magnetic field of $3$~G and an artificial cut-off value on the minimum temperature to ensure numerical stability. In this work, we performed simulations with the mixed heating model, depending on both $B$ and $\rho$ at the same time (model \texttt{M}), as in \cite{Mok2008}. \cite{KY2017} also briefly discuss a 3D simulation with anti-shearing motions and $(\alpha,\beta) = (1,1)$, although their main result focuses on an $(\alpha,\beta) = (2,0)$ configuration. In full agreement with the results of \cite{KY2015}, model \texttt{M} only leads to condensations in flux ropes formed through anti-shearing motions. This can be explained by the increased magnetic energy stored in the field by shearing motions, which act to amplify $B_z$ (Fig.~\ref{fig:all_magnetic}) and hence increase the local heating rate within the flux rope. Moreover, regions throughout the simulation domain where the magnetic pressure is high, like the converging loop footpoints and the flux rope edges, also experience a high heating rate that leads to unfavourable conditions for the condensation process. Hence, for those simulations involving anti-shearing motions, only the bottom of the flux rope and lower atmosphere with increased density could provide sufficient radiative losses to initiate the thermal instability and lead to the formation of one large monolithic condensation in the flux rope and a string of coronal rain blobs in the surrounding atmosphere. Then, as already mentioned, our adoption of a $10$~G background field strength, although bringing us closer to observational values \citep{Casini:2003, Wang:2020}, increases the core heating rate even further.

% The high footpoint heating could perhaps be overcome by modifying the heating rate within the bottom layer of domain.

%  As before, this can be explained by the nature of the background heating, i.e., enhanced heating at the lateral edges of the flux rope due to compression of the field lines on the one hand, and a decreasing $B_z$ component and hence heating rate on the other hand:   \nb{Connection to `hot nuggets' in coronal cavities? Found some references, but none of them were \textit{really} comparable to simulations here.} \jen{Indeed, they aren't really related, but from a conversation with the referee that we had we figured this was the nearest comparison that could be made given the extremely limited selection of observational case studies. We don't have to make a direct connection here though, since it's not the focus of the paper. We can of course add something later if the referee asks once again.}

As mentioned above, attempts to form condensations with the combined use of shearing motions and $B$-dependent heating models had thus far proven unsuccessful. Here, we introduced the physically motivated dynamic masked heating reduction throughout the flux rope interior, and this is shown to lead to successful prominence formation.
Indeed, as anti-shearing motions are not commonly observed along PILs \citep[][where differential rotation usually increases the shear of magnetic arcades]{gibb2014shear,mackay2015shear}, it was previously highly puzzling as to how a prominence could form under `typical' solar conditions since 2.5D models indicated such behaviour would categorically prohibit prominence formation when adopting $B$-dependent background heating. Moreover, the spontaneous coronal rain phenomenon reported here indicates that anti-shearing motions may decrease magnetic energy too substantially to consistently explain prominence formation.
%\RK{It is worth highlighting once more how this behaviour is likely heavily influenced by the adopted 2.5D domain and associated boundary conditions. In fact, the 3D study of %\cite{Xia2016a} already shows that ample condensations develop within a flux rope formed via positive shearing motions and with an exponential heating prescription, although this is also true for our 2.5D simulations. And so, a future study that combines our reduced heating approach with either a larger 2.5D domain such as in %\cite{Zhao2017}, or considers a fully 3D domain as in %\cite{KY2017}, but for the case of positive shearing motions, is likely to address or even overcome many of the shortcomings identified within this and previous studies.}{}

\subsection{(Additional) influence of Joule heating} \label{sec:joule_heating}

For all of the setups presented here, we adopted a constant $\eta$ value for the resistivity throughout the simulation space and time. This permits a range of resistive evolutions within the simulations, such as magnetic reconnection (but see also Section~\ref{sec:additional_results}), with perhaps the most ubiquitous of them being the dissipation of strong currents as Joule heating $\eta |\textbf{j}|^2$. In Table~\ref{tab:joule_heating}, we quantify the magnitude of this Joule heating at two typical locations, and compare it against the contribution from the assumed background heating model.

\begin{table}[]
    \centering
    \begin{tabular}{l|cccc}
        \hline
        \hline
        Heating & \multicolumn{4}{c}{Joule} \\
        (erg~cm$^{-3}$~s$^{-1}$) & \multicolumn{4}{c}{\underline{Background}} \\[0.7ex]
        \hline
        Model & \texttt{E} & \texttt{RE} & \texttt{M1} & \texttt{RM1} \\
        \hline
        FR centre & 8.5e-5 & 8.3e-5 & 7.5e-5 & 8.5e-5 \\
        \phantom{---}(final snapshot)    & \underline{6.3e-4} & \underline{1.9e-4} & \underline{5.4e-4} & \underline{2.8e-4} \\[0.7ex]
        Reconnection sites & 1.4e-1 & 1.4e-1 & 2.4e-1 & 1.1e-1 \\
        \phantom{---}(t = 1888 s) & \underline{1.1e-3} & \underline{1.1e-3} & \underline{1.5e-3} & \underline{1.1e-3} \\
        \hline
    \end{tabular}
    \caption{Compared magnitude of Joule and background heating for all heating models. The former is usually smaller or comparable to the latter, except for locations of reconnection and mass slippage, i.e. strong localised currents.}
    \label{tab:joule_heating}
\end{table}

In the centre of the flux rope at the end of the simulation, we find the Joule heating contribution to be of the same order of magnitude, but in all cases smaller than the contribution of the background heating model. During the  preceding formation process, however, the picture is quite the opposite with the Joule heating contributing a considerable amount of energy to the local plasma \citep[cf. the alternating hot shells in][]{jack2020}, most significantly at the footpoint locations during their migration towards the PIL.
% ; from Eq.~\ref{eq:MHD_full_magnetic} we can see how the motions here lead to an enhancement in $\textbf{j}$, the energy of which is subsequently directly dissipated into the local plasma, and additionally so during the nearby reconnection events. \jen{Since $J = \nabla \times B$, I wonder how much, if at all, this is influenced by the divB cleaning approaches - just a thought}. 

It is clear that the evolutions within the simulations lead to the generation of strong currents. In all cases, the influence that these currents have on the thermodynamic properties are directly proportional to the magnitude of the assumed resistivity $\eta$ since the Joule heating scales linearly with it. In the current simulation, we deliberately overestimated the magnitude of this resistivity so as to assist in the flux rope formation process - a magnitude that was necessarily above the numerical resistivity limit set by the resolution. In future simulations we may relax this overestimation, as an increase in spatial resolution will similarly decrease the restriction on the resolvable resistivity. Subsequent current enhancements may peak to higher values with smaller extents, depositing the equivalent energy, but we nevertheless suggest this leading to an overall reduction in the amplitude of the associated Joule heating across the simulation domain. The non-linear nature of these simulations requires, however, dedicated studies so as to accurately conclude the influence of these modifications. For more dynamic scenarios, one may also explore the influence of the spatially varying `anomalous' prescriptions, as in \citet{Zhao2017} and \citet{Ruan2020}.

\subsection{Towards increased realism in prominence simulations} \label{sec:towards_realism}

Despite the ever-increasing body of work on the coronal heating problem, no single heating mechanism or combination thereof has yet been deemed the definitive solution \citep[remaining key questions are highlighted by][]{klimchuk2015}. How this heating affects prominence formation is here shown to differ between various parametric models. Specifically, the timescale on which instability occurs, decreases for parametric models, as seen on Fig.~\ref{fig:comparison_models_densities}, which also indicates that imposing a time-independent exponential heating background without differentiating between the flux rope and the surrounding atmosphere imposes a non-negligible residual heating rate inside the flux rope, affecting the onset of thermal instability: it was shown that applying a heating reduction inside the flux rope decreases this time significantly. Then, for models \texttt{(R)M1}, we found the directional choice of shearing motions to have a large impact, explaining why condensations appear at a slightly later time for the simulation that was paired with positive shearing despite employing the reduced heating model. It is thus expected that the same reduced heating setup with anti-shearing motions would instead lead to a considerably earlier development of condensations compared to the standard \texttt{M1} setup. 

In Table~\ref{tab:heating_comparison}, we list the $n_{\mathrm{max}}$ and $T_{\mathrm{min}}$ characteristics of the prominences towards the end of the simulations, as formed by the different heating models. Despite the different models employed, the maximum number density and minimum temperature are comparable for most heating prescriptions, indicating that the physics governing the prominence material, once formed, is independent of the heating prescriptions explored here. Furthermore, prominence observations summarised by \cite{Parenti2014} list number densities ranging from $10^{9} - 10^{11}$~cm$^{-3}$, which shows that the values obtained in all our simulations are in good agreement. The same may be said for the core temperature of the prominences of our simulations, which is reported by the same author to have a range of $7500 - 9000$~K. The spatial dimensions of our simulated prominences are, however, on the smaller side of observed widths between $1-10$~Mm resulting from the relatively small computational domain. Pressures inside the prominence are found in the range $0.08 - 0.41$~dyn/cm$^2$ for models \texttt{E} and \texttt{M1} and in the range $0.05 - 0.64$~dyn/cm$^2$ for their reduced counterparts, in perfect correspondence with observations \citep{Parenti2014}.

\begin{figure}
	\centering
	\resizebox{\hsize}{!}{\includegraphics{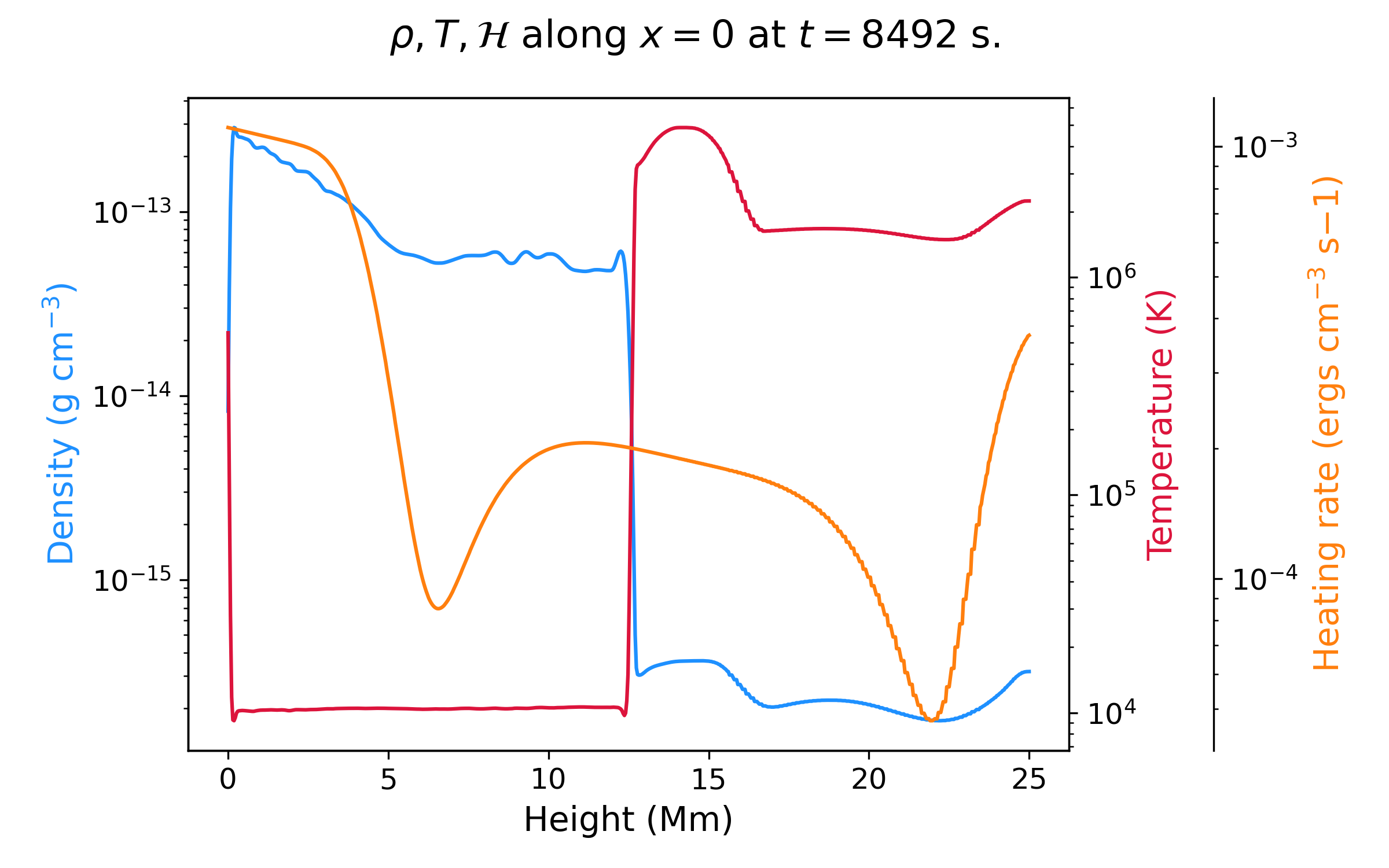}}
	\caption{Vertical profile of density, temperature and heating rate through the prominence for model \texttt{RE}. The density varies exponentially in the lower part of the prominence and is more uniform in the upper part inside the heating mask.}
	\label{fig:profile_rho_bQ}
\end{figure}

Panel~a of Fig.~\ref{fig:comparison_models_area_mass} clearly shows evaporation of prominence material for models (\texttt{R})\texttt{E}, which is found to occur at the prominence top where the cool material resides next to the hot flux rope core.
This is possibly a result of Joule heating through the large perpendicular currents, which could provide a net heating effect in the neighbouring parts of the prominence.
One distinct difference arises between the average prominence densities of models with and without heating reduction in Fig.~\ref{fig:comparison_models_densities}: the former have lower average densities. We calculated the distributions of prominence density and pressure to seek an explanation. Models \texttt{RE}, \texttt{RM1} produce very peaked distributions around the (lower) average values, indicating that the resulting prominences are largely uniform in pressure and density. The uniform region takes up most of the prominence, which does show stratification at top and bottom ends. For model \texttt{M1}, the distributions have a larger standard deviation but are still rather peaked, although around multiple higher density or pressure values. For model \texttt{E}, the distributions are much broader, pointing to prominences with a strongly pronounced internal density and pressure stratification, which clearly appears on a vertical profile taken within the prominence. This stratification can be solely attributed to the steady exponential heating background, which imposes a vertical stratification inside the prominence body instead of a stratification along field lines as would be expected for plasma bound to the magnetic field \citep{Blokland:2011a, Blokland:2011b}. Indeed, a similar stratification is observed in the bottom part of the prominence that escapes the modified heating mask of model \texttt{RE} -- a comparison between Figs.~\ref{fig:reduced_heating} and \ref{fig:mix_red_evolution} indicates how the lowest-most extent of the prominence lies outside the ellipse. In particular, the part of the prominence for model \texttt{RE} inside the reduced heating mask -- which has almost uniform density and pressure -- is seen to extend further horizontally in the magnetic dips compared to the bottom end due to the different thermodynamic properties of the plasma, more specifically due to a different pressure scale height. A vertical cut along the prominence, featured on Fig. \ref{fig:profile_rho_bQ}, clearly shows an exponential density profile in the prominence material outside the reduced heating mask, and almost uniform conditions above the location where the heating mask is applied. Such a feature, in particular, provides a particularly strong argument as to why an exponential background should not be applied indiscriminately to 2.5D prominence formation in flux ropes: not only does the exponential influence the timescales, but also the stratification of plasma properties within the flux rope - a key component in synthesising any resulting simulation as an observation \citep[e.g.][]{Jenkins:2022}.
% \nb{Is the uniformity then related to assumptions/observations in $H\alpha$ spectroscopy?} \jen{Yes! Whether that's in turn connected to reality is another question but it's something that we can comment on i.e., the fact they are in agreement but that we have artificially influenced this with a reduced internal heating which may or may not be \textit{realistic}.} 
%\nb{Some comment on relation to $H\alpha$-observations?}

% Although the exact prescription of the heating model has a clear influence on the details and potential suppression of prominence formation through levitation-condensation, the resulting prominences are comparable in terms of number density and temperature, as was shown in Table~\ref{tab:heating_comparison}. Moreover, we already indicated \nb{or maybe better do this here?} that these values are also in good agreement with observations \citep{Parenti2014}, so that creating prominences with realistic masses should be a matter of increasing the computational domains in the future. 
To compare our obtained prominence masses against observations, we multiply the mass per unit length by a typical flux rope or prominence length of $100$ Mm -- this assumption agrees with the $\delta$ parameter used in the reduced heating prescription \eqref{eq:reduced_heating_length}, although the prominence plasma rarely extends down towards the chromospheric footpoints of a flux rope and this estimation hence gives an upper limit. The estimated prominence masses then lie between $2.3 - 8.3 \times 10^{13}$~g, a similar order of magnitude to previous simulation results \citep{jack2020}, but still below the typical inferred values from observations of $10^{14} - 2 \times 10^{15}$~g \citep{Parenti2014}. Hence, the inclusion of more-realistic considerations for the heating, however valid, do not address the outstanding issue of creating more `realistic' prominences in terms of order-of-magnitude total mass content. Still, the reduced heating approach does increase the prominence masses compared to the other heating models. The larger masses can generally be attributed to the increased net energy loss rate due to the decrease in background heating, which in turn makes a larger portion of the flux rope more prone to thermal instability. The correspondence between observational and simulation number densities and temperatures, yet the persistent underestimation for total mass content indicates that the simulation domain and subsequent dimensions of the ab initio prominences are smaller than present within the actual solar atmosphere. The consideration of both a larger domain, and one that includes a chromosphere, would provide the coronal flux rope with an abundance of additional material, as already shown with the realistic prominence mass values found in the work of \cite{Zhao2017}. Moreover, the justification for the artificial heating reduction employed in our 2.5D simulations is automatically accounted for in 3D simulations, since field-aligned thermal conduction is unable to efficiently transport energy towards the middle of the flux rope. We thus envision the combination of a 3D setup as in \cite{KY2017} that implements localised heating considerations and drives the formation of a flux rope via positive shearing motions with the presence of a chromosphere, to be the natural next step. Moreover, the reduced heating model presented in this work based on common assumptions in 1D prominence formation models can benefit from recent progress, for example by \cite{huang2021}, who combined evaporation and injection into one model. Extending our setup as described above could additionally incorporate these effects in the current levitation-condensation model by supplying localised heating in the chromosphere. It is anticipated that the above aspects will be crucial in both further advancing these models towards self-consistency, and facilitating effective comparisons against observations. The results obtained in this paper hence will, and should, one day be superseded by more realistic models of a complete solar atmosphere.

% \jen{Also need to connect each of these statements to the need for fully 3D representations to overcome the limitations of the 2.5D representations. The 2.5D approaches are, of course, justified from a pure plasma physics perspective i.e., exploration of the CCI, TI, etc, under extreme resolutions. But, of course, the evaporation-condensation mechanism is likely more useful to explain \textit{actual} prominences. It's completely okay to concede this in the paper - perhaps Rony will rephrase it to sound less negative, who knows.}

\subsection{Additional results: slippage and oscillations} \label{sec:additional_results}

As reported in earlier work with the same setup, the high resistivity value in our simulations enhances the effect of mass slippage over the field lines \citep{jack2020,low2012a}. Due to this effect, the prominence body as a whole performs motions perpendicular to the poloidal magnetic field, where through resistive dissipation it sinks towards the bottom of the simulation domain. We observe this phenomenon in all our simulations, with its typical signature of increased current density at the slippage sites. In a benchmark simulation, where resistivity was switched off after the formation of the flux rope, this slippage effect was all but eliminated.

The observed mass slippage also affects the prominence oscillations reported in this work. Condensations form at different heights throughout the flux rope on distinct poloidal field lines. After falling down, they perform oscillatory motions about $x=0$ with different periods, as is clearly visible in Fig~\ref{fig:exp_red_evolution}. The periodicity is governed by the pendulum model, with dependence on the local radius of curvature of the magnetic field \citep{Arregui2018}. The oscillations found in this work occur spontaneously in response to evolving condensation formation and do not require an imposed perturbation. Longitudinal oscillations have been invoked to explain counter-streaming motions in prominences \citep{zirker1998}, as they have been found in other simulations of prominence formation \citep[e.g.][]{Xia2011}. 
An analysis of the oscillations along selected field lines -- using the method from \cite{liakh2020} -- for model \texttt{RE} indeed reveals a period dependence on height, where the period increases from top of the prominence (region closest to the centre of the host flux rope) to bottom with decreasing the radius of field curvature \citep[the accompanying analysis is included in a recent conference proceeding,][]{brughmansflux}. In particular, oscillation damping times also show a height dependence. Oscillations are even temporarily amplified inside the top regions of the prominence as well as in its bottom part, as was reported for the first time by \cite{liakh2020} and \cite{liakh2021}. Again from the benchmark simulation without resistivity, we find that in general, mass slippage decreases the apparent periods and increases the apparent damping time. We note that the oscillations observed here are plane-projected manifestations of longitudinal oscillations. Observations of longitudinal prominence oscillations yield typical periods of $50-60$ minutes \citep{Arregui2018}, somewhat longer than the periods of $5-10$ minutes reported in our work, which again underlines the need to consider larger simulation domains and full 3D models with line-tied flux rope ends in future works.

\subsection{Condensation velocities and phase space}

An analysis of the maximum velocity over the domain provides further insight in the formation and evolution of condensations. During the formation phase, inflow velocities of $55 - 100$~km/s are found. These velocities decrease after the formation of the initial condensation for models \texttt{M1}, \texttt{RM1}, but for the exponential models, subsequently, velocities up to $190$~km/s are observed within those falling condensations that originate near the apex of the flux rope. These high velocities, in excess of those anticipated from free fall acceleration, are a consequence of pressure evolution associated with the formation of condensations at other locations along the same flux surface. This surplus of kinetic energy driven by dramatic pressure evolution within the exponential models is the reason why the material performs damped oscillations with large initial amplitudes after formation.

The phase space evolution in Fig.~\ref{fig:phase_space} featured a non-isobaric evolution of the thermal instability for model \texttt{RE}. Models \texttt{E}, \texttt{M1}, and \texttt{RM1} also lead to a very similar evolution, although in the very onset of thermal instability, the distribution follows the pressure isocontours a bit longer than for model \texttt{RE}. The non-isobaric evolution of the condensation process does not come as a surprise: the inclusion of gravity introduces non-isobaric dynamics -- \cite{jack2020} already found baroclinicity localised to the condensations. In short, the onset of thermal instability visually matches isobaric conditions, but not far into the linear evolution, the isochoric criterion becomes more appropriate as the temperature decreases faster than the density increases \citep{Xia2012}. \cite{moschou2015} also arrived at this conclusion by identifying the dominant instability criterion in the thermally unstable regions associated with coronal rain.

% Conclusions

\section{Conclusions} \label{sec:conclusions}
We performed 2.5D simulations of prominence formation through levitation-condensation with two classes of heating models, and an additional ad hoc modification for 2.5D simulations that approximates the 3D flux rope structure, and tracks the flux rope during runtime based on magnetic curvature. Exponential heating models lead to larger prominences but have a high residual heating rate inside the flux rope and prominence, while the mixed models have the more realistic assumptions but do not produce condensations when combined with positive shearing motions. Both of these inconsistencies are overcome by the reduced heating mask: the reduced exponential model leads to large condensations, almost at the lower edge of observed masses for whole prominences, and simultaneously eliminates the residual flux rope heating, thus creating a much cooler flux rope and prominences with almost uniform density and pressure. Crucially, those models with both magnetic-field-strength-dependent heating and the ad hoc reduction mask, combined with shearing motions, now lead to the successful formation of condensations. As both classes of heating models are shown to lead to prominences with different evolutions and morphologies but very similar physical properties, taking these models to either a larger 2D or even fully 3D domain will likely overcome many of the problems and inconsistencies that continue to reside within our simulations, and in turn further increase the simulated prominence masses. Finally, a phase-space visualisation of the condensation process describes neither isobaric nor isochoric behaviour, but rather a combination of the two, with a state of constant pressure along flux surfaces recovered once force-balance is achieved within the flux rope. 

A natural progression of this work is to extend the setup to 3D, using a larger simulation domain with the inclusion of a chromosphere. This will enable us to assess the effect of the mixed heating model on a flux rope formed through positive shearing motions and whether the resulting prominences can reach realistic masses. Indeed, the inclusion of a chromosphere and transition region will drastically modify the energy balance of the model atmosphere, which will have a large effect on stability provided by thermal conduction of energy towards the transition region, where it is radiated away. The results of this work, which only considers a coronal domain, could then be modified significantly in view of this added realism, where perhaps thermal instability is less likely to occur except when evaporation from the chromosphere is taken into account. In fact, such a setup has been successfully implemented by \cite{Xia2016a}. It remains unclear how the cooling of the longest field lines obtained by \cite{KY2017} would be affected by the presence of a chromosphere and transition region. 

% Furthermore, associated oscillations due to prominence formation will be truly longitudinal in 3D and it will be interesting to compare these to observations. 

% Additionally, as \cite{KY2015} point out, shearing motions oftentimes lead to the activation and subsequent eruption of the flux rope, a process for which recent models DO SOMETHING NEW \citep{Fan2017}.

% \nb{Additional extensions to our flux rope tracking mechanism could be improved robustness in dealing with flux rope mergers, or incorporate tracking of and heating reduction in secondary flux ropes, even evolving the reduction factor throughout the simulation to account for the changing flux rope length during formation.} 

% \RK{}{RK: check the references, keep those that matter (now it is a very long list! You can AT MOST have 20 pages with everything included!!!}

\begin{acknowledgements}
NB acknowledges support by his Fonds Wetenschappelijk Onderzoek (FWO) fellowship (grant number 11J2622N) and would like to thank Valeriia Liakh for her advice on analysing prominence oscillations. JMJ and RK received funding from the European Research Council (ERC) under the European Unions Horizon 2020 research and innovation programme (grant agreement No. 833251 PROMINENT ERC-ADG 2018), and from Internal Funds KU Leuven, project C14/19/089 TRACESpace. The computational resources and services used in this work were provided by the VSC (Flemish Supercomputer Center), funded by the Research Foundation Flanders (FWO) and the Flemish Government - department EWI. Simulation visualisations were created using \texttt{yt} (\hyperlink{https://yt-project.org/}{https://yt-project.org/}).
\end{acknowledgements}

% Bibliography
\bibliographystyle{apalike}
\bibliography{Thesis_bibliography.bib}

\begin{appendix}

\section{Flux rope tracking}\label{sec:flux_rope_tracking}
%\textit{Explain how and why algorithm works (on its own!), arbitrary choices.}

We here set forth to describe the algorithm used to track the flux rope during runtime. First, the 3D field curvature is calculated over the domain using,
\begin{equation} \label{eq:curvature}
    \boldsymbol{\kappa} = \mathbf{b}\cdot \nabla \mathbf{b},
\end{equation}
where $\mathbf{b} = \frac{\mathbf{B}}{B}$ is the unit vector along the magnetic field. 
% Indeed, if $\mathbf{r}(s)$ is the arc-length parameterisation of the magnetic field, then basic differential geometry defines the curvature as $\boldsymbol{\kappa} = \frac{d\mathbf{b}}{ds}$. Using the chain rule, this expression becomes $\frac{d\mathbf{r}}{ds}\cdot \nabla\mathbf{b}$. As $\mathbf{b}$ is the unit vector tangent to the field, we arrive at the expression \eqref{eq:curvature}.
Since there is no variation in the $z$-direction in our 2.5D setup, the gradient in \eqref{eq:curvature} has only non-trivial components along $x$ and $y$. The centre of the flux rope is then detected as a point of minimal curvature as it corresponds to a straight field line that locally is oriented directly out-of-plane in the 2.5D representation. As we do not deliberately displace the central axis of the flux rope in our simulations \citep[cf.][]{jack2020}, we can assume that the flux rope centre lies directly above the PIL and search for a point $(0,y_O)$. This detection method is in accordance with \cite{curvature}, who found that weak curvature correlates with a strong normal force (i.e. straight field lines in 2.5D) and strong curvature occurs in the neighbourhood of points with a weak magnetic field (i.e. 3D O-points). The flux rope edges are detected instead as local maxima in the curvature, resulting from the dynamical evolution of the flux rope in the surrounding atmosphere.

% The centre of the flux rope may also be defined as an O-point of the poloidal field, for which alternative detection methods exist \citep[e.g.,][]{O_point_detection}. In particular,  O- or X-points (locations of reconnection) are found as critical points of the magnetic flux function $A$ whose nature is determined by eigenvalues of the Hessian (maxima are O-points, saddle points are X-points). The flux rope can then be defined as the region where $A$ takes on values between the maximum value at the flux rope centre and the value reached at the X-point at the bottom of the flux rope \citep[see, e.g.,][]{Zhao2017}. For this work, we adopt the more computationally trivial curvature prescription.

% Without giving the full algorithm, we now present the flux rope tracking strategy. 
\subsection{Detailed algorithm for tracking}~\\

Suppose an $N \times N$ grid with resolution $\Delta x \times \Delta y$ is given by a collection of points $\{ (x_i,y_i) \ | \ i = 1, \ldots, N \}$. The characteristics of the model ellipse are fixed by three values: let $y_O$ be the height of the flux rope centre and $x_b = x_O+b,\ y_a = y_O+a$ the horizontal and vertical coordinate of the corresponding edges, ascertained at the previous simulation time. The new locations in the current timestep are then determined by solving optimisation problems of the locations and curvature, which are solved in a discrete manner by looping over the AMR grids. 
% Since our simulations employ parallel computing and AMR, this means one must first determine the optimum over all grids allocated to a single processor, after which the optimum over all processors can be found. These optimisation problems are solved in a discrete manner by looping over all grids.

To find the flux rope centre, we solve the following optimisation problem, 
\begin{align*}
% 	& \textit{FR centre detection:} \\[1.ex]
    & \text{Maximise } y_i \\[.7ex]
    & \begin{array}{cl}
        \text{such that} & | x_i | \leq \Delta x, \\[0.7ex]
                           & \kappa (x_i,y_i) \text{ is minimal and }\leq 0.01, \\[0.7ex] 
                           & |y_i - y_O| < d, \\[0.7ex]
                           & 0 < i < N.
    \end{array}
\end{align*}
If no new location of minimal curvature satisfying these conditions is found, the maximal allowed jump $d$ is increased by $10\%$ and the problem solved again, up to a maximum of $d = 0.2$. Here, $d$ varies throughout the simulation to accommodate the more dynamic periods of flux rope formation and prevent `overshooting'. Its value in code units depends on the variable timestep $\Delta t$ and is given by
\begin{equation} \label{eq:d_optimisation}
    d = \begin{cases} 1 & \text{if } y_O, x_b, y_a \text{ have not all been found,} \\
    10 \Delta t & \text{if } y_O, x_b, y_a \text{ have all been found.} \end{cases}
\end{equation}
% The above optimisation problem may hence be reformulated as `finding a local minimum in curvature close to zero in a strip of two cells around $x=0$, that is not located on an edge of the grid and should lie within a distance of $d$ from the previous location'.

After the coordinate $y_O$ has been found, $b$ is found by looking for maxima of $\kappa$ over a horizontal strip at height $y_O$ while $a$ is found in the same way over a vertical strip at $x_O=0$. To find $b$, the following problem is solved,
\begin{align*}
    % & \textit{Horizontal edge detection:} \\[1.ex]
    & \text{Maximise } x_i \\[.7ex]
    & \begin{array}{cl}
        \text{such that} & | y_i-y_O | \leq \Delta y, \\[0.7ex]
                           & x_i \geq 0, \\[0.7ex]
                           & \kappa (x_i,y_i) \text{ is maximal}, \\[0.7ex] 
                           & |x_i - x_b| < 5d, \\[0.7ex]
                           & 0 < i < N.
    \end{array}
\end{align*}
If the above problem has no solution, $d$ is again increased in steps of $10\%$ until it reaches $d=0.1$. While approximately $d<45\Delta t$, we first look for solutions $x_i \geq x_b$ that increase the dimensions of the flux rope since it is expanding for most of its dynamic evolution, but this condition is relaxed when no solution is found. 
% The optimisation problem amounts to `finding a local maximum in curvature in a horizontal strip of two cells around $y=y_O$, that is not located on an edge of the grid and should lie within a distance of $5d$ from the previous location'. 
The optimisation problem for finding the location of the vertical edge $y_a$ of the flux rope is similar, with $x_{i,b}$ interchanged with $y_{i,a}$. 
% If for any of the above three optimisation problems no solution is reached, the condition that excludes grid edge points is relaxed.

\subsection{Flux rope tracking results}

% \begin{figure}
% 	\centering
%	\begin{subfigure}{0.49\textwidth}
%		\resizebox{\hsize}{!}{\includegraphics{Figures_reduced/reduced_FR_top.png}	}
%	\end{subfigure}
%	\begin{subfigure}{0.49\textwidth}
%		\resizebox{\hsize}{!}{\includegraphics[width=\textwidth]{Figures_reduced/reduced_b.png}}	
%	\end{subfigure}
% 	\begin{subfigure}{\hsize}
% 		\resizebox{\hsize}{!}{\includegraphics{Figures/reduced_FR_center.png}}
% 	\end{subfigure}
% 	\begin{subfigure}{\hsize}
% 		\resizebox{\hsize}{!}{\includegraphics{Figures/reduced_flux.png}}
% 	\end{subfigure}
% 	\caption{The total flux in the flux rope is approximated by $B_z \pi a b$, where the perpendicular magnetic field component is taken to be the maximum over the flux rope. The jumps around $7000$/$7500$ s are a result of a jump in the horizontal axis length $b$.}
% 	\label{fig:FR_tracking}
% \end{figure}

The tracked location of the flux rope centre is shown to oscillate as a result of mergers between the primary and subsequent subordinate flux ropes \citep[cf.][]{jack2020}. We find an average initial period of $280 - 300$~s. 
% For the simulation with model \texttt{RE}, this increases to $320$~s towards the end of the simulation. 
% With model \texttt{RM1}, the flux rope centre is pushed upwards by the very high temperatures in the outer shell after $\sim$~5500~s.

The magnetic flux through the tracked flux rope is approximated as $\Phi = B_z \pi a b$, where $B_z$ is taken to be the field strength reached in the flux rope centre. The obtained values are hence rather an order-of-magnitude estimate, more so since the horizontal length $b$ is underestimated after the end of the flux rope formation. 
% The magnetic flux evolution between $2000 - 3000$~s also shows how the detection of the horizontal edge suffers from the mergers of secondary flux ropes. 
% The large jumps in magnetic flux around $7000$~s for model \texttt{RE} and around $7500$~s for model \texttt{RM1} are a result of the detection of a new local maximum in curvature along the horizontal axis, and hence a new value for $b$. In reality, the flux rope is steady from approximately $3000$~s. 
We find the estimated magnetic flux bound by the flux rope, across models, to be of order $10^{19}$~Mx, consistent with earlier simulations \citep[e.g.][]{Zhao2017} yet several orders of magnitude below observational results \citep[e.g. 10$^{21}$~Mx in][]{Dissauer:2018} which leads to the conclusion that we are working with a small flux rope in these simulations.

The tracked flux rope is not always in a 1-to-1 correspondence with the actual flux rope. First, the tracking algorithm can suffer from mis-identifications whenever the field topology undergoes a sudden evolution during, for instance, flux rope mergers, but the shape is recovered after a few time units. Second, the assumption of a perfectly elliptic poloidal field will by definition exclude some portions of the flux rope immediately above the X-point as the topology is deformed by a combination of magnetic pressure and tension associated with the ongoing reconnection \citep[cf.][]{jack2020}. Third, once the footpoint driving motions are switched off the flux rope is assumed to be fully formed. However, due to the large value of $\eta$ maintained here on out, a slower diffusion of field lines leads to the expansion of the flux rope partially out of the upper domain boundary. After this time, the previous methods and justifications for defining the edge of the flux rope, through curvature features alone, are no longer applicable. To avoid inconsistencies and mis-identifications, we opt to fix the horizontal half-length $b$ from the moment the driving velocities cease, and update $b$ if and only if a higher value is detected. Hence, as time progresses, the detected dimensions tend to underestimate the actual size of the flux rope. An example of this underestimation can be observed in Fig.~\ref{fig:reduced_heating}, panel d.

%The magnetic flux is seen to lie in the range $2 - 3 \times 10^{11}$ Wb as the flux rope reaches its maximal size. The sudden jump towards the end of the simulation results from an updated value for the minor axis $b$ of the ellipse. \cite{Zhao2017} found magnetic fluxes of the same order, going up to $10^{12}$. BUT BETTER COMPARE THIS TO ACTUALLY OBSERVED VALUES. The qualitative evolution of the flux rope area is very similar to the total flux since $B_z$ does not vary significantly after the flux rope has formed. The total flux rope area in our simulation reaches $250$ Mm$^2$, which takes up more than $40\%$ of the simulation domain. In reality, the flux fills much of the domain, which shows that the boundary conditions really limit the size of the flux rope.

\end{appendix}

\end{document}